\documentclass[lettersize,journal]{IEEEtran}
\usepackage{amsmath,amsfonts}
\usepackage{algorithmic}
\usepackage{algorithm}
\usepackage{array}
\usepackage[caption=false,font=normalsize,labelfont=sf,textfont=sf]{subfig}
\usepackage{textcomp}
\usepackage{stfloats}
\usepackage{url}
\usepackage{verbatim}
\usepackage{graphicx}
\usepackage{cite}

\usepackage{newtxtext,newtxmath}
\usepackage{xcolor}
\usepackage{soul}
\usepackage{multirow}
\usepackage{booktabs}
\usepackage{tabularx}
\usepackage{enumitem}
\usepackage{comment}

\usepackage{float}
\usepackage[sort&compress]{cleveref}
    \crefname{equation}{}{}
    \Crefname{equation}{}{}
    \crefformat{equation}{(#2#1#3)}
    \Crefformat{equation}{(#2#1#3)}
    \crefrangeformat{equation}{(#3#1#4)--(#5#2#6)}
    \Crefrangeformat{equation}{(#3#1#4)--(#5#2#6)}
    \crefrangemultiformat{equation}
      {(#3#1#4)--(#5#2#6)}
      {, (#3#1#4)--(#5#2#6)}
      {, (#3#1#4)--(#5#2#6)}
      {, (#3#1#4)--(#5#2#6)}
    \Crefrangemultiformat{equation}
      {(#3#1#4)--(#5#2#6)}
      {, (#3#1#4)--(#5#2#6)}
      {, (#3#1#4)--(#5#2#6)}
      {, (#3#1#4)--(#5#2#6)}

\newcommand{\case}[1]{\textbf{#1}}

\makeatletter
\newcommand{\STAGE}[1]{\item[] {\bfseries\itshape #1}}
\makeatother
\begin{document}
\IEEEaftertitletext{\vspace{-1.2\baselineskip}}

\raggedbottom

\bstctlcite{BSTcontrol}

\title{Stochastic Power-Water Coordination: Unlocking Flexibility in Hybrid RO Desalination Plants via Variable-Speed Pumps and Tank Mixing}
\author{%
Rongxing Hu,~\IEEEmembership{Member,~IEEE}, 
\and
Charalambos Konstantinou,~\IEEEmembership{Senior Member,~IEEE}%
\thanks{R. Hu and C. Konstantinou are with the Computer, Electrical and Mathematical Sciences and Engineering (CEMSE) Division,
King Abdullah University of Science and Technology (KAUST), Thuwal 23955, Saudi Arabia.(e-mail: rongxing.hu@kaust.edu.sa; charalambos.konstantinou@kaust.edu.sa)}
}



\maketitle

\begin{abstract}
Water desalination plants (DPs) are among the most critical infrastructures and largest electricity loads in water-scarce regions worldwide. Although reverse osmosis (RO) desalination is the most energy-efficient and dominant technology, it remains energy-intensive but can offer substantial flexibility potential for power systems. This paper proposes a coordinated operation framework for power systems and DPs that explicitly accounts for both systems’ operational constraints and fully unlocks DP flexibility. To achieve this, a detailed DP model is developed, incorporating the characteristics of an actual high-pressure pump with variable-speed operation, on-off operation with flushing requirements, water quality constraints, and water dynamics and salt mixing in the storage tank. By proactively managing freshwater storage and tank salinity in a closed-loop coordinated scheduling framework, the operational flexibility of the DP is significantly enhanced. With appropriate simplification and linearization, the resulting coordinated scheduling problem is formulated as a tractable mixed-integer linear programming (MILP) model, and a two-step decomposed commitment-scheduling stochastic optimization (TDCSO) is proposed to efficiently address uncertainties. Case studies validate the proposed approach and demonstrate up to a 6\% operating cost reduction. 
\end{abstract}

\vspace{-0.3\baselineskip}

\section{Introduction}
\IEEEPARstart{G}{lobally}, the rapid growth of renewable energy sources (e.g., PV and wind) necessitates sufficient flexibility resources to address the inherent intermittence and variability of renewable generation and maintain reliable power-system operation, including grid-level energy storage and large-scale demand response. In parallel, freshwater supply in islands and arid/semi-arid regions relies heavily on energy-intensive desalination plants (DPs), which constitute large electricity loads. In Saudi Arabia, desalination supplied over 70\% of drinking water in 2020 and is projected to reach about 90\% by 2030~\cite{bg-2024-towards}, while desalination is expected to represent around 15\% of total energy consumption in the Middle East by 2040~\cite{bg-IEA-2019-Desalinated}. Reverse osmosis (RO) is the dominant desalination technology due to its high energy efficiency, accounting for more than 90\% of newly installed capacity~\cite{bg-UN-2022-Desalination}. With appropriate design or retrofitting, RO desalination plants (RO-DPs) can provide responsive flexibility by participating in demand response and ancillary service markets~\cite{bg-ERCOT-2016-Desalination}, making coordinated RO-DP and power-system operation an effective means to improve renewable integration, reduce costs, and enhance system reliability~\cite{bg-res-2016-water}.

A hybrid desalination plant (HDP) co-located with renewables can benefit seawater RO (SWRO) desalination by reducing costs and carbon emissions~\cite{bg-res-2016-water}. Many studies have demonstrated that RO-DPs can provide flexibility for renewable integration. Using an experimental SWRO plant,~\cite{cap-2004-RO-7000} evaluates flexibility for wind integration by varying pressure and flow through pump speed control using a variable-frequency drive (VFD), with over 7,000~hrs of operation showing no membrane degradation. A detailed dynamic RO model is developed in~\cite{cap-2016-modeling} to assess ramping capability, response time, and load-following performance under variable renewable generation, demonstrating that RO desalination can operate as a flexible load. Operational flexibility can be further enhanced by grouping RO modules with adjustable flow and pressure~\cite{cap-2013-optimal}. An MPC-based operation strategy is proposed in~\cite{cap-2022-variable} to adapt RO operating parameters and setpoints to renewable fluctuations and is validated using a laboratory-scale test rig.

Several recent studies have investigated coordination between RO-DPs and power systems; however, the desalination process is complex, due to nonlinear high-pressure pump (HPP) characteristics~\cite{pump-2008-modeling}, highly nonconvex RO process behavior~\cite{book-2022-desalination}, and water storage dynamics, so most works rely on simplified DP models. In~\cite{dp-2024-operation}, coordinated day-ahead operation of power-water operation is studied considering DC power flow, while the DP is represented only through operational bounds. In~\cite{dp-2019-coordinated,dp-2022-evaluate,dp-2024-flexibility}, desalination is modeled via a simplified ``power-freshwater flow'' curve under a constant recovery rate:~\cite{dp-2019-coordinated} co-optimizes bulk power systems and DPs and quantify operational cost reductions, ~\cite{dp-2022-evaluate} assesses the virtual energy-storage capability of DP-equipped water systems in small communities,~\cite{dp-2024-flexibility} embeds this model in an adaptive robust optimization framework to quantify DP flexibility, showing up to 20\% cost reduction. Similarly,~\cite{dp-2021-optimal-energy} derives a ``freshwater flow-power'' relationship for constant-speed pump-driven DPs in microgrids. Recognizing significant RO process efficiency losses caused by frequent off-optimal operation under day-ahead demand response,~\cite{dp-2019-a-novel} further incorporates both ``freshwater flow-power” and ``efficiency-power” relationships of DPs and proposes using efficiency thresholds within a two-stage scheduling framework.

The aforementioned works~\cite{dp-2024-operation,dp-2019-coordinated,dp-2022-evaluate,dp-2024-flexibility,dp-2021-optimal-energy,dp-2019-a-novel} rely on simplified equivalent DP models and therefore overlook detailed characteristics of HPPs, RO desalination processes, and freshwater storage tanks. To address this limitation, several recent studies explicitly model DP subcomponents~\cite{dp-2023-coordinating,utah-2020-optimal-WDN,utah-2019-optimal-desali,kfupm-2022-risk}. In~\cite{dp-2023-coordinating}, a day-ahead scheduling problem is formulated for a power system with DPs and hybrid energy storage; however, the HPP is modeled with constant efficiency via a simplified hydraulic-energy expression (feed flow × feed pressure with coefficients), neglecting detailed RO desalination dynamics. In~\cite{utah-2020-optimal-WDN} and~\cite{utah-2019-optimal-desali}, coordinated power-water operation with DPs is studied, with the latter further considering participation in frequency regulation markets. These works capture the ``head-flow-speed'' and ``power-flow-speed'' characteristics of variable-speed pumps, enabling a wider operational range than constant-speed pumps, but adopt a generic pump formulation for both HPPs and water distribution pumps despite their distinct operational requirements. The RO process is modeled only as a function of membrane pressure differences and freshwater production, without representing the detailed desalination process. In~\cite{kfupm-2022-risk}, a two-stage stochastic co-optimization framework is proposed that considers temperature effects, concentration polarization~\cite{cp-2005-RO}, and salt transport in the RO process, enabling explicit calculation of water quality (e.g., salinity or total dissolved solids (TDS)). Nevertheless, the HPP model remains simplified, capturing only hydraulic energy with constant efficiency, and assuming a constant recovery rate, which ultimately limits achievable DP flexibility~\cite{cap-2022-variable}.

Overall, existing works rely on simplified or generic DP models that capture only partial operational characteristics, and water quality regulation is largely overlooked except in~\cite{kfupm-2022-risk}. Essential operational requirements, such as membrane flushing after shutdown and before restart, which consumes non-negligible water and energy to protect costly RO membranes~\cite{book-2021-RO-flush}, and pretreatment stabilization prior to restart~\cite{stanford-2024-pretreat-RO}, are also typically omitted. In addition, end-of-horizon storage-TDS regulation is not considered, which can cause salt accumulation and deteriorate long-term operation. Moreover, although various simplification and linearization approaches have been proposed (e.g., shrinking the feasibility region or neglecting constraints), flexibility estimates are highly sensitive to model accuracy, and most studies lack validation against laboratory experiments or detailed numerical DP models. Last but not least, DP operation affects not only freshwater quantity in storage tanks but also salt content, quantified by TDS and directly linked to end-user water quality; however, existing studies mainly exploit flexibility from the desalination process and freshwater storage while ignoring TDS dynamics. Inspired by~\cite{2019-active-salinity}, which demonstrates that appropriately mixing low- and high-TDS sources can reduce the levelized cost of water in the CAISO market, this work treats storage TDS as a flexibility resource by regulating stored-water TDS, rather than strictly constraining instantaneous freshwater-generation TDS. The underlying rationale is to reduce power consumption (higher TDS) during high-price or high net-load periods and increase power consumption (lower TDS) when electricity prices are low or renewable generation is abundant. To the best of our knowledge, no existing work has proactively managed storage TDS as a flexibility resource for coordinated water-power system operation. In light of the above, the main contributions of this paper are threefold:

\textit{(1)} A coordinated DP-power system scheduling framework is developed using a detailed DP model with manufacturer-specified variable-speed HPP characteristics and detailed RO dynamics, including flushing consumption, water-quality constraints, tank TDS mixing, and water/salt balances. With introduced simplifications and linearization, the original nonconvex mixed-integer nonlinear programming problem is converted into a tractable mixed-integer linear programming (MILP) formulation suitable for operational and planning studies.

\textit{(2)} Stored-water TDS flexibility is exploited by regulating tank TDS while meeting end-user water-quality requirements, thereby closing the loop on tank water quality over the scheduling horizon and significantly expanding flexibility. To incorporate uncertainty while maintaining tractability, we propose a two-step decomposed commitment-scheduling stochastic optimization (TDCSO) algorithm that separates commitment decisions from stochastic flexibility scheduling.

\textit{(3)} The proposed approach is validated against the original full desalination process model, and extensive case studies  numerically demonstrate its effectiveness and advantages. End-of-horizon storage requirements and sensitivities to water demand and operating pressure are further analyzed to provide practical implementation insights.

The rest of the paper is organized as follows: Section~II introduces the HDP and power system component models and operational constraints. Section~III formulates the coordinated HDP-power system problem. Finally, Section~IV presents the results, while Section~V concludes the paper.

\begin{figure}[!ht]
    \vspace{-6pt}
    \centering
    \includegraphics[width=0.9\linewidth]{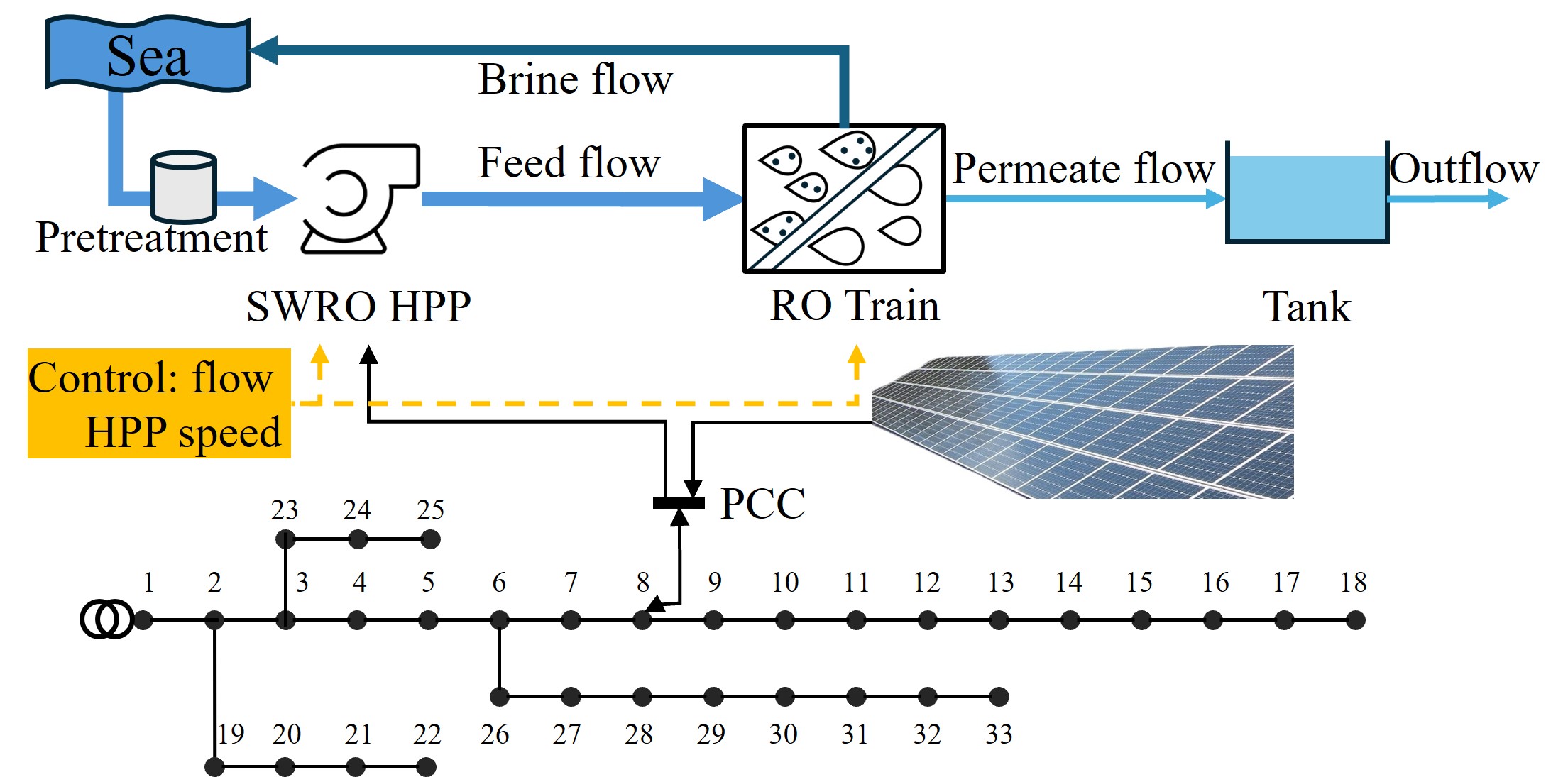}
    \caption{Diagram of the HDP and the power distribution system.}
    \label{fig:sys_layout}
    \vspace{-6pt}
\end{figure}

\section{Modeling the HDP and Power System}
A typical SWRO HDP is shown in Fig.~\ref{fig:sys_layout}. It includes an HPP, RO trains consisting of multiple parallel pressure vessels equipped with spiral-wound RO membrane elements, the permeate/freshwater storage tank, and a co-located PV system to reduce energy costs and carbon emissions. Pretreated seawater is pumped into the RO trains under high pressure, where membrane separation produces a high-TDS brine flow and a low-TDS permeate flow. In addition, as a large load in the power system, the SWRO HDP is required not to cause regulation issues in the power system network, like low voltage and high line loss. Moreover, since water demand varies significantly and the renewables are intermittent, the desalination process should be able to turn on/off and adjust its operation points to adapt to those changes. The flexibility of DPs is achieved by the variable-speed operation of the HPP (controlling flow and pressure) and the modular operation of RO units (controlling flow)~\cite{cap-2022-variable}.

\begin{figure}[t]
    \vspace{-8pt}
    \centering
    \includegraphics[width=1\linewidth]{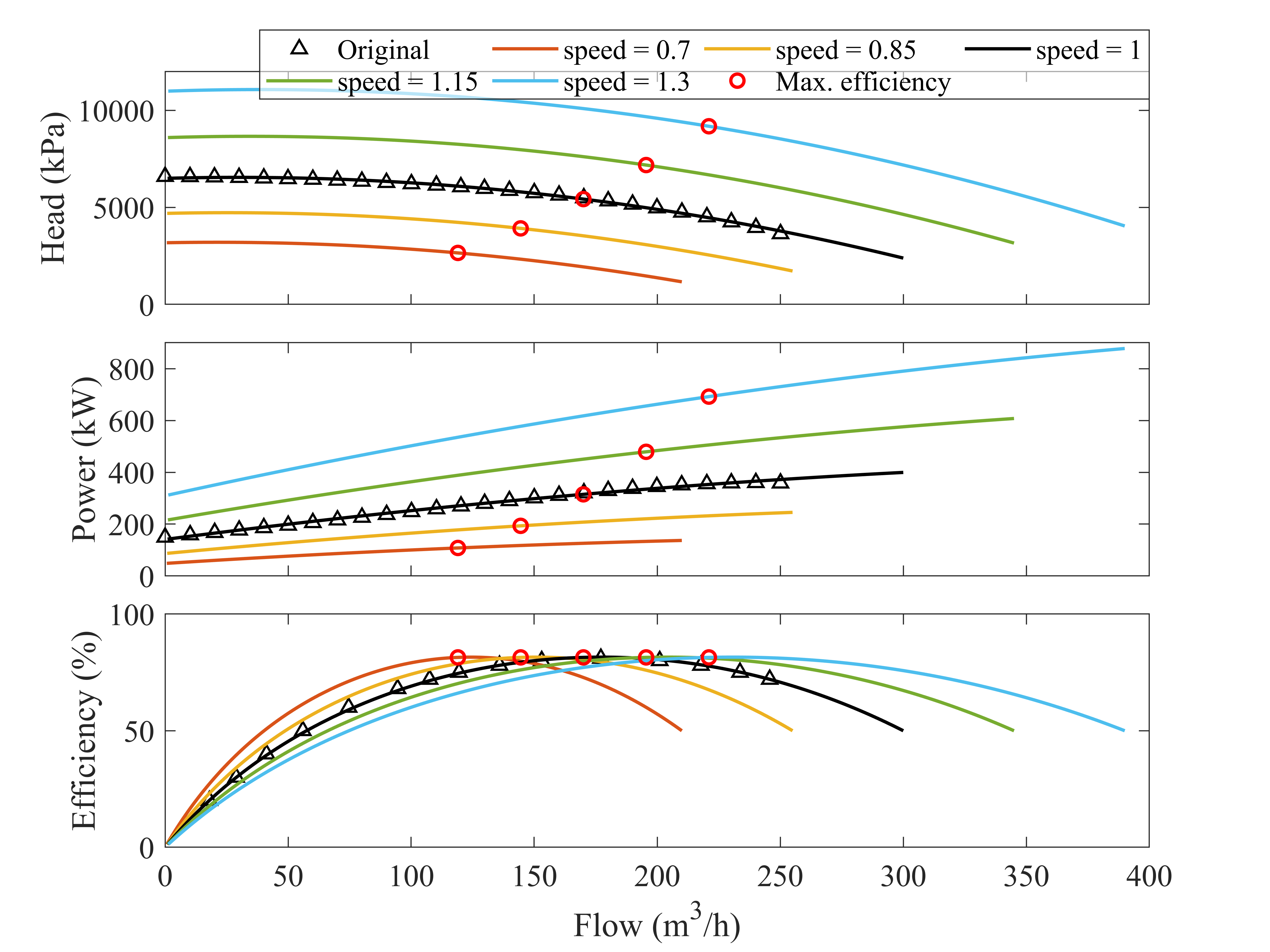}
    \caption{Characteristic curves of a VFD-enabled variable-speed HPP. Black triangles denote the original data at nominal speed in the product manual, and red circles indicate the operating points with maximum efficiency.}
    \label{fig:hpp}
    \vspace{-6pt}
\end{figure}

\vspace{-2mm}
\subsection{High Pressure Pump System}
The HPP accounts for most of the SWRO desalination energy consumption. Because the SWRO process is primarily governed by pump operation, the available flexibility is determined by the HPP’s feasible operating range. In this work, we consider a VFD-enabled variable-speed HPP (VSP), which can operate at different pump speeds and therefore offers a wider operational range while maintaining high efficiency. 

Due to the high head requirement (up to 83 bar), SWRO plants typically use multistage centrifugal HPPs~\cite{pump-2001-multistage-RO}, whose developed head/pressure at a given flow rate scales with the number of stages. Rather than adopting simplified hydraulic models~\cite{kfupm-2022-risk} or generic pump models~\cite{utah-2020-optimal-WDN,utah-2019-optimal-desali}, we develop a VFD-enabled HPP model based on a real multistage centrifugal SWRO HPP product, the KSB Multitec 100~\cite{pump-ksb-multitec}. It should be noted that the product manual provides characteristic curves (head-flow, power-flow) only at nominal speed, while off-nominal operation at variable speed typically follows the affinity laws~\cite{book-2008-pump-affinity} (see \eqref{eq:hpp_affinity}), which describe how operating points change with pump speed and are used here to model variable-speed HPP operation. As shown in Fig.~\ref{fig:hpp}, increasing the pump speed increases the achievable flow/pressure, maximum flow, and power consumption. Moreover, pump efficiency degrades significantly when the operating point deviates from the best-efficiency point at a given speed. Hence, using the hydraulic flow-head product while ignoring characteristic curves (Fig.~\ref{fig:hpp}) or assuming constant pump efficiency can cause substantial errors unless only small operating-point variations are allowed.
\begin{equation}
    \frac{F_{\omega_2}^{\text{hpp}}}{F_{\omega_1}^{\text{hpp}}}
    = \frac{\omega_2}{\omega_1}, \quad
    \frac{H_{\omega_2}^{\text{hpp}}}{H_{\omega_1}^{\text{hpp}}}
    = \left( \frac{\omega_2}{\omega_1} \right)^2, \quad
    \frac{P_{\omega_2}^{\text{hpp}}}{P_{\omega_1}^{\text{hpp}}}
    = \left( \frac{\omega_2}{\omega_1} \right)^3
    \label{eq:hpp_affinity}
\end{equation}
where $\omega_1$ and $\omega_2$ denote two pump speeds; 
$F_{\omega_1}^{\text{hpp}}$, $H_{\omega_1}^{\text{hpp}}$, $P_{\omega_1}^{\text{hpp}}$, 
$F_{\omega_2}^{\text{hpp}}$, $H_{\omega_2}^{\text{hpp}}$, and $P_{\omega_2}^{\text{hpp}}$ 
represent the corresponding water flow rate, water head, and HPP power consumption, respectively.

The pump model consists of the pump characteristic functions and the operation constraints. The feed head/pressure and the HPP power consumption are calculated by \eqref{eq:hpp_head}-\eqref{eq:hpp_power}. Equations \eqref{eq:hpp_flow_limit}-\eqref{eq:hpp_power_limit} denote the operational bounds on the feed flow, feed head, HPP speed, and HPP power, which are determined by the pump characteristics. Please note that, unlike in water distribution systems,  dramatic and large variations in the seawater feed head can damage the RO membranes, therefore, the feed head is regulated within a narrow range, and RO desalination is generally considered not suitable for fast frequency regulation. The active/reactive power of the high-pressure pump system (HPS), including the VFD, motor, and HPP, is calculated by \eqref{eq:hps_power}.
\begin{equation}
    H_t^{\text{fd}} =
    N_{\text{stg}}^{\text{hpp}}\left( a_2 (F_t^{\text{fd}})^2
    + a_1 F_t^{\text{fd}} \omega_t + a_0 \omega_t^2 \right)
    \label{eq:hpp_head}
\end{equation}
\begin{equation}
    P_t^{\text{hpp}} =
    N_{\text{stg}}^{\text{hpp}}\left( b_2 (F_t^{\text{fd}})^2 \omega_t
    + b_1 F_t^{\text{fd}} \omega_t^2 + b_0 \omega_t^3 \right)
    \label{eq:hpp_power}
\end{equation}
\begin{equation}
    0 \le F_t^{\text{fd}} \le F_{\max}^{\text{hpp}} \, \omega_t
    \label{eq:hpp_flow_limit}
\end{equation}
\begin{equation}
    H_{\min}^{\text{fd}} U_t^{\text{on}} \le H_t^{\text{fd}} \le H_{\max}^{\text{fd}} U_t^{\text{on}}
    \label{eq:hpp_head_limit}
\end{equation}
\begin{equation}
    \omega_{\min} U_t^{\text{on}}
    \le \omega_t
    \le \omega_{\max} U_t^{\text{on}}
    \label{eq:hpp_speed_limit}
\end{equation}
\begin{equation}
    0 \le P_t^{\text{hpp}} \le P_{\max}^{\text{hpp}} U_t^{\text{on}}
    \label{eq:hpp_power_limit}
\end{equation}
\begin{equation}
    P_t^{\text{hps}} = \frac{P_t^{\text{hpp}}}{\eta^{\text{mt}} \eta^{\text{vfd}}}, \;
    Q_t^{\text{hps}} = k_{\text{QP}}^{\text{hps}} P_t^{\text{hps}}
    \label{eq:hps_power}
\end{equation}
where $F_t^{\text{fd}}$, $H_t^{\text{fd}}$, and $P_t^{\text{hpp}}$ are the feed flow, feed pressure, and HPP power. The parameters $a_2$, $a_1$, $a_0$ and $b_2$, $b_1$, $b_0$ are the corresponding coefficients, and $N_{\text{stg}}^{\text{hpp}}$ denotes the number of stages of the multistage centrifugal HPP. $\omega_t$ denotes the normalized pump speed, with allowable range $\omega_{\min}$ to $\omega_{\max}$. The binary variable $U_t^{\text{on}}$ represents the desalination on/off status. $F_{\max}^{\text{hpp}}$ is the maximum HPP flow at nominal speed. $H_{\min}^{\text{fd}}$ and $H_{\max}^{\text{fd}}$ are the lower and upper bounds of the feed pressure/head. $P_{\max}^{\text{hpp}}$ is the maximum pump power set by the motor rating. $P_t^{\text{hps}}$ and $Q_t^{\text{hps}}$ are the HPS active/reactive powers; $k_{\text{QP}}^{\text{hps}}$ is the coefficient relating reactive and active power, and $\eta^{\text{mt}}$ and $\eta^{\text{vfd}}$ are the motor and VFD efficiencies, respectively.

\vspace{-2mm}
\subsection{RO Desalination Process}
The RO desalination process includes coupled water and salt mass-balance equations, subject to RO membrane specifications and the feed flow and pressure. We adopt an equivalent pressure vessel-level model~\cite{manual-2010-filmtec-RO}. 

\textit{1) Water balance:} The water balance is enforced by \eqref{eq:flow_balance}. Equation \eqref{eq:permeate_flow} describes how the RO membrane permeability, pressure across the membranes, and osmotic pressure determine the permeate water production, while \eqref{eq:ro_pressure_feasible} ensures non-negative permeate flow. The RO recovery rate (permeate-to-feed flow ratio) is bounded by \eqref{eq:recovery_ratio}. Constraints \eqref{eq:off_flow_limit}-\eqref{eq:br_fd_relation} specify the permissible flow ranges based on the operating status of the desalination process and the capabilities of the RO system.
\begin{equation}
    F_t^{\text{fd}} = F_t^{\text{br}} + F_t^{\text{pe}}
    \label{eq:flow_balance}
\end{equation}
\begin{equation}
    F_t^{\text{pe}} = k_W^{\text{ro}} \left( \Delta H_t^{\text{ro}} - \Delta \pi_t^{\text{ro}} \right)
    \label{eq:permeate_flow}
\end{equation}
\begin{equation}
    \Delta H_t^{\text{ro}} \ge \Delta \pi_t^{\text{ro}}
    \label{eq:ro_pressure_feasible}
\end{equation}
\begin{equation}
    F_t^{\text{fd}} R_{\min}^{\text{rec}}
    \le F_t^{\text{pe}} \le F_t^{\text{fd}} R_{\max}^{\text{rec}}
    \label{eq:recovery_ratio}
\end{equation}
\begin{equation}
    F_{\min}^{\text{fd}} U_t^{\text{on}} 
    \le F_t^{\text{fd}} \le F_{\max}^{\text{fd}} U_t^{\text{on}}
    \label{eq:fd_flow_limit}
\end{equation}
\begin{equation}
    0 \le F_t^{\text{pe}}, F_t^{\text{br}} \le F_{\max}^{\text{fd}} U_t^{\text{on}}
    \label{eq:off_flow_limit}
\end{equation}
\begin{equation}
    0 \le F_t^{\text{br}} \le F_t^{\text{fd}}
    \label{eq:br_fd_relation}
\end{equation}
where $F_t^{\text{br}}$ and $F_t^{\text{pe}}$ are the brine and permeate flows. $k_W^{\text{ro}}$ is the permeate-production coefficient, $k_W^{\text{ro}}=K_W^{\text{mb}}A^{\text{mb}}N^{\text{mb}}T_t^{\text{mb}}$, where $K_W^{\text{mb}}$, $A^{\text{mb}}$, $N^{\text{mb}}$, and $T_t^{\text{mb}}$ are the water permeability coefficient, RO membrane area, number of RO membrane elements, and temperature correction factor~\cite{book-2022-desalination}, respectively. $\Delta H_t^{\text{ro}}$ and $\Delta \pi_t^{\text{ro}}$ are the average pressure difference and osmotic pressure difference across the membrane. $F_{\min}^{\text{fd}}$ and $F_{\max}^{\text{fd}}$ bound the feed flow, and $R_{\min}^{\text{rec}}$ and $R_{\max}^{\text{rec}}$ bound the recovery rate.

\textit{2) Pressure balance:} The pressure across the membranes is calculated in \eqref{eq:ro_head_diff} using the average concentrate-side pressure and the permeate pressure. The brine pressure is obtained from \eqref{eq:brine_head}, which accounts for frictional pressure losses along the membranes, while \eqref{eq:pe_head_limit} and \eqref{eq:pe_head_set} specify the permeate water pressure/head. The osmotic pressure across the membranes is given by \eqref{eq:ro_osmotic_diff}--\eqref{eq:ro_pressure_condition}. The osmotic pressures of the flows are determined by their TDS levels, as described in \eqref{eq:osmotic_fd_br}--\eqref{eq:osmotic_perm}.

\vspace{-3mm}
\begin{equation}
    \Delta H_t^{\text{ro}} = (H_t^{\text{fd}} + H_t^{\text{br}})/2 - H_t^{\text{pe}}
    \label{eq:ro_head_diff}
\end{equation}
\begin{equation}
    H_t^{\text{br}} = k_{\text{H}}^{\text{fric}} H_t^{\text{fd}}
    \label{eq:brine_head}
\end{equation}
\begin{equation}
    H_t^{\text{pe}} \le H_t^{\text{br}}
    \label{eq:pe_head_limit}
\end{equation}
\begin{equation}
    H_t^{\text{pe}} = H_{\text{set}}^{\text{pe}} U_t^{\text{on}}
    \label{eq:pe_head_set}
\end{equation}
\begin{equation}
    \Delta \pi_t^{\text{ro}}
    = C^{\text{cp}}\!\left(\pi_t^{\text{fd}} + \pi_t^{\text{br}}\right)/2 - \pi_t^{\text{pe}}
    \label{eq:ro_osmotic_diff}
\end{equation}
\begin{equation}
    0 \le {\Delta \pi}_t^{\text{ro}} \le \Delta H_t^{\text{ro}}  
    \label{eq:ro_pressure_condition}
\end{equation}
\begin{equation}
    \pi_t^{\text{fd}} = k^{\text{os}} S_t^{\text{fd}} U_t^{\text{on}}, \;
    \pi_t^{\text{br}} = k^{\text{os}} S_t^{\text{br}}
    \label{eq:osmotic_fd_br}
\end{equation}
\begin{equation}
    \pi_t^{\text{pe}} = k^{\text{os}} S_t^{\text{pe}}
    \label{eq:osmotic_perm}
\end{equation}
where $\Delta H_t^{\text{ro}}$ is the pressure difference across the RO membrane; $H_t^{\text{br}}$ and $H_t^{\text{pe}}$ are the brine- and permeate-side pressures; $k_{\text{H}}^{\text{fric}}$ is the friction coefficient; and $H_{\text{set}}^{\text{pe}}$ is the permeate pressure. $\Delta \pi_t^{\text{ro}}$ is the osmotic pressure difference across the RO membrane, with $\pi_t^{\text{fd}}$, $\pi_t^{\text{br}}$, and $\pi_t^{\text{pe}}$ denoting the feed-, brine-, and permeate-side osmotic pressures. $S_t^{\text{fd}}$, $S_t^{\text{br}}$, and $S_t^{\text{pe}}$ are the corresponding TDS levels, and $k^{\text{os}}$ is the osmotic-pressure coefficient. $C^{\text{cp}}$ is the concentration polarization factor~\cite{book-2022-desalination}.

\textit{3) Salt Balance:} Salt transport and separation in the desalination process determine the permeate water quality. The salt balance is ensured by  \eqref{eq:salt_balance}. As shown in \eqref{eq:perm_tds}, the permeate salt mass is governed by the RO membrane salt permeability and the average TDS of the feed and brine flows calculated in \eqref{eq:ro_tds_ave}. The brine and permeate TDS levels are bounded by \eqref{eq:brine_tds_limit} and \eqref{eq:perm_tds_limit}.
\begin{equation}
    S^{\text{fd}} F_t^{\text{fd}}
    = S_t^{\text{br}} F_t^{\text{br}} + S_t^{\text{pe}} F_t^{\text{pe}}
    \label{eq:salt_balance}
\end{equation}
\begin{equation}
    S_t^{\text{pe}} F_t^{\text{pe}}
    = k_S^{\text{ro}} \left( C^{\text{cp}} S_t^{\text{ro}} - S_t^{\text{pe}} \right)
    \label{eq:perm_tds}
\end{equation}
\begin{equation}
    S_t^{\text{ro}} =
    \frac{ S^{\text{fd}} F_t^{\text{fd}}
    + S_t^{\text{br}} F_t^{\text{br}}}{ F_t^{\text{fd}} + F_t^{\text{br}}}
    \label{eq:ro_tds_ave}
\end{equation}
\begin{equation}
    S^{\text{fd}} U_t^{\text{on}} 
    \le S_t^{\text{br}} \le S_{\max}^{\text{br}} U_t^{\text{on}}
    \label{eq:brine_tds_limit}
\end{equation}
\begin{equation}
    0 \le S_t^{\text{pe}} \le S_{\max}^{\text{pe}} U_t^{\text{on}}
    \label{eq:perm_tds_limit}
\end{equation}
where $k_{S}^{\text{ro}}$ is the RO salt-transport coefficient, $k_{S}^{\text{ro}} = K_{S}^{\text{mb}} A^{\text{mb}} N^{\text{mb}} T_{t}^{\text{mb}}$, and $K_{S}^{\text{mb}}$ is the membrane salt-permeability coefficient. $S_{t}^{\text{ro}}$ represents the average concentrate-side TDS computed from the feed and brine flows. $S_{\max}^{\text{br}}$ and $S_{\max}^{\text{pe}}$ denote the maximum allowable brine and permeate TDS, respectively.

\begin{figure}[!htbp]
    \vspace{-5pt}
    \centering
    \includegraphics[width=1.0\linewidth]{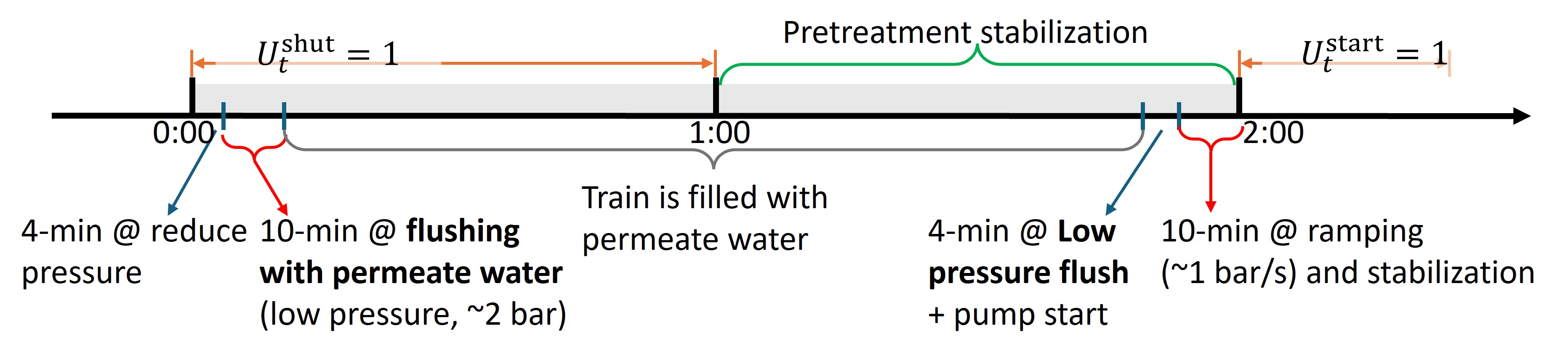}
    \caption{Flushing requirement of an example RO-DP.}
    \label{fig:flush}
    \vspace{-12pt}
\end{figure}

\textit{4) Flushing Requirements:} When the desalination process is shut down for a few hours or longer, the RO trains must be flushed and cleaned after shutdown and during restart to prevent membrane damage caused by fouling, scaling, and biological growth~\cite{manual-2010-filmtec-RO}. This work focuses on day-ahead operation with off periods shorter than one day, and demand response services require loads to return to operation within a few hours, for example, Responsive Reserve Service in ERCOT requires recovery within 3~hrs~\cite{bg-ERCOT-2016-Desalination}. In such cases, low-pressure flushing (typically 10--30~min, including ramping) using permeate water is required, and the RO trains are kept filled with permeate water while idle. As illustrated in Fig.~\ref{fig:flush}, the flushing procedure consumes both water and energy, including HPP ramp-down/ramp-up energy consumption, as well as the required permeate water. However, if the shutdown lasts for several days or longer, additional cleaning steps, such as applying preservation solutions, are necessary.
Equations~\eqref{eq:flush_water}--\eqref{eq:flush_energy} calculate flushing water and energy consumption,
~\eqref{eq:shutdown_logic_1}--\eqref{eq:shutdown_logic_2} determine the hours in which shutdown actions occur, while 
\eqref{eq:start_logic_1}--\eqref{eq:start_logic_2} identify the hours when the system has just fully restarted. 
Equation~\eqref{eq:shutdown_min} ensures the minimum off duration required for pretreatment stabilization~\cite{stanford-2024-pretreat-RO}.
\begin{equation}
    W_t^{\text{flush}}
    = W_{\text{use}}^{\text{shut}} U_t^{\text{shut}}
    + W_{\text{use}}^{\text{start}} U_{t+1}^{\text{start}}
    \label{eq:flush_water}
\end{equation}
\begin{equation}
    E_t^{\text{flush}}
    = E_{\text{use}}^{\text{shut}} U_t^{\text{shut}}
    + E_{\text{use}}^{\text{start}} U_{t+1}^{\text{start}}
    \label{eq:flush_energy}
\end{equation}
\begin{equation}
    U_{t-1}^{\text{on}} - U_t^{\text{on}}
    \le U_t^{\text{shut}}
    \le U_{t-1}^{\text{on}}
    \label{eq:shutdown_logic_1}
\end{equation}
\begin{equation}
    U_t^{\text{shut}} \le 1 - U_t^{\text{on}}
    \label{eq:shutdown_logic_2}
\end{equation}
\begin{equation}
    U_t^{\text{on}} - U_{t-1}^{\text{on}}
    \le U_t^{\text{start}}
    \le U_t^{\text{on}}
    \label{eq:start_logic_1}
\end{equation}
\begin{equation}
    U_t^{\text{start}} \le 1 - U_{t-1}^{\text{on}}
    \label{eq:start_logic_2}
\end{equation}
\begin{equation}
    \sum\nolimits_{z=t}^{t+D^{\text{shut}}-1} (1-U_z^{\text{on}})
    \;\ge\; D^{\text{shut}}\,U_t^{\text{shut}}
    \label{eq:shutdown_min}
\end{equation}
where $W_t^{\text{flush}}$ and $E_t^{\text{flush}}$ are flushing water and energy use. $U_t^{\text{shut}}$ and $U_t^{\text{start}}$ indicate shutdown and just-fully-restarted hours. $W_{\text{use}}^{\text{shut}}$ and $W_{\text{use}}^{\text{start}}$ ($E_{\text{use}}^{\text{shut}}$ and $E_{\text{use}}^{\text{start}}$) are the water (energy) used during shutdown and restart. $D^{\text{shut}}$ is the minimum off duration required for pretreatment stabilization.

\subsection{Storage Tank}
The water storage tank is typically much cheaper than the battery energy storage systems (BESSs), its capacity can support one to several days water reserve. The water storage tank enables the desalination process to exploit its flexibility and function as a virtual BESS (e.g., load shifting). However, the dynamics of the water storage tank involves two coupled forms of storage, water and salt, due to the mixing of flows with different TDSs. Although user-end water quality is directly determined by the tank water quality (TDS), existing works consider only the water dynamics and overlook the salt/TDS dynamics. The tank water balance is given by ~\eqref{eq:tk_balance}. Constraint ~\eqref{eq:tk_capacity} enforces the storage operating range while satisfying reserve requirements, and ~\eqref{eq:tk_terminal} ensures tend-of-horizon storage is no less than the initial level.
\begin{equation}
    W_t^{\text{tk}} =
    W_{t-1}^{\text{tk}}
    + F_t^{\text{pe}} \Delta t
    - F_t^{\text{out}} \Delta t
    - W_t^{\text{flush}}
    \label{eq:tk_balance}
\end{equation}
\begin{equation}
    W_{\min}^{\text{tk}}
    \le W_t^{\text{tk}}
    \le W_{\max}^{\text{tk}}
    \label{eq:tk_capacity}
\end{equation}
\begin{equation}
    W_T^{\text{tk}} \ge W_{\text{ini}}^{\text{tk}}
    \label{eq:tk_terminal}
\end{equation}
where $W_t^{\text{tk}}$ is the tank water volume at time $t$ and $W_{\text{ini}}^{\text{tk}}$ is the initial storage. $W_{\min}^{\text{tk}}$ and $W_{\max}^{\text{tk}}$ bound the operational storage range. $F_t^{\text{out}}$ denotes the tank outflow (water consumption).

Tank TDS mixing calculations are necessary to track stored-water quality and enable proactive TDS management, and the mixing time can be neglected~\cite{cap-2013-optimal}. The tank salt balance in~\eqref{eq:tank_salt_balance} accounts for salt in existing tank water, permeate inflow, outflow, and flushing water. Because the flushing volume is relatively small and tank-TDS variations are typically limited under proper regulation, the flushing-water TDS is approximated by an estimated value $S_{\text{ave}}^{\text{tk}}$, a practical way is to use the average tank TDS from previous days. Similarly, since hourly water usage is small relative to tank storage capacity (which may cover several days),~\eqref{eq:tank_out_tds} approximates the outflow TDS using the average tank TDS over two consecutive hours. In addition, constraint \eqref{eq:tank_out_tds_limit} regulates tank and outflow water quality to satisfy end-user requirements. Finally,~\eqref{eq:tank_end_tds} ensures that the end-of-horizon tank TDS does not exceed its initial value, avoiding long-term deterioration caused by salt accumulation in tanks.
\begin{equation}
    S_t^{\text{tk}} W_t^{\text{tk}}
    = S_{t-1}^{\text{tk}} W_{t-1}^{\text{tk}}
    + S_t^{\text{pe}} F_t^{\text{pe}} \Delta t
    - S_t^{\text{out}} F_t^{\text{out}} \Delta t
    - S_{\text{ave}}^{\text{tk}} W_t^{\text{flush}}
    \label{eq:tank_salt_balance}
\end{equation}
\begin{equation}
    S_t^{\text{out}} = (S_t^{\text{tk}} + S_{t-1}^{\text{tk}})/2
    \label{eq:tank_out_tds}
\end{equation}
\begin{equation}
    0 \le S_t^{\text{tk}}, S_t^{\text{out}} \le S_{\max}^{\text{out}}
    \label{eq:tank_out_tds_limit}
\end{equation}
\begin{equation}
    S_T^{\text{tk}} \le S_{\text{ini}}^{\text{tk}}
    \label{eq:tank_end_tds}
\end{equation}
where $S_t^{\text{tk}}$ is the tank-water TDS at time $t$ and $S_{\text{ini}}^{\text{tk}}$ is its initial value. $S_t^{\text{out}}$ is the outflow TDS, $S_{\max}^{\text{out}}$ is the maximum allowable end-user TDS. $S_{\text{ave}}^{\text{tk}}$ is the estimated flushing water TDS.

Instead of strictly constraining the instantaneous freshwater-generation TDS $S_{\max}^{\text{pe}}$, this work exploits TDS flexibility in freshwater storage tanks by regulating the stored-water TDS $S_{\max}^{\text{out}}$, which directly determines delivered water quality. This permits $S_{\max}^{\text{pe}} > S_{\max}^{\text{out}}$, thereby further unlocking TDS-mixing flexibility in the RO process beyond the conventional water-volume storage flexibility. To the best of our knowledge, this is the first work to do so.

\vspace{-2mm}
\subsection{Co-located PV}
The PV operation is bounded by following constraints:
\begin{equation}
    0 \le P_t^{\text{pv}} \le P_t^{\text{pvPred}}, \;
    0 \le Q_t^{\text{pv}} \le P_{\text{rate}}^{\text{pv}}
    \label{eq:pv_power_limits}
\end{equation}
\begin{equation}
    P_t^{\text{pv}} - \sqrt{2}\, P_{\text{rate}}^{\text{pv}}
    \le Q_t^{\text{pv}} \le
    - P_t^{\text{pv}} + \sqrt{2}\, P_{\text{rate}}^{\text{pv}}
    \label{eq:pv_inverter_limit}
\end{equation}
where $P_t^{\text{pvPred}}$ is the predicted PV output; $P_t^{\text{pv}}$ and $Q_t^{\text{pv}}$ are the utilized PV active/reactive powers; $P_{\text{rate}}^{\text{pv}}$ is the PV inverter rating.

\vspace{-2mm}
\subsection{Power Distribution Network}
The LinDistFlow~\cite{math-2002-distflow-33} is used to calculate the power flow, ensuring proper coordination between the SWRO desalination system and the power network and preventing violations of distribution system limits, such as low voltage and line overloading. The power balance is enforced by~\eqref{eq:node_power_balance}, and the net load is computed by~\eqref{eq:hdp_power_injection}--\eqref{eq:load_bus_active}. The corresponding reactive power equations are omitted. Line-loading limits are given in~\eqref{eq:line_rating_box}--\eqref{eq:line_rating_lower}. Node voltages are computed by \eqref{eq:voltage_drop} and limited by~\eqref{eq:voltage_limit}, with substation voltage regulated by \eqref{eq:substation_voltage}.

\vspace{-2mm}
\begin{equation}
    \sum_{i \in \Omega_j^{\text{from}}}
    \!\!\begin{bmatrix} P_{ij,t} \\ Q_{ij,t} \end{bmatrix}
    =
    \sum_{k \in \Omega_j^{\text{to}}}
    \!\!\begin{bmatrix} P_{jk,t} \\ Q_{jk,t} \end{bmatrix}
    +
    \begin{bmatrix} P_{j,t} \\ Q_{j,t} \end{bmatrix}
    \label{eq:node_power_balance}
\end{equation}
\begin{equation}
    P_{j,t} = P_t^{\text{hdp}} + P_t^{\text{load}}
    \quad j = j^{\text{hdp}}
    \label{eq:hdp_power_injection}
\end{equation}
\begin{equation}
    P_t^{\text{hdp}}
    = P_t^{\text{hps}} - P_t^{\text{pv}}
    + {E_t^{\text{flush}}}/{\Delta t}
    \label{eq:hdp_power_coupling}
\end{equation}
\begin{equation}
    P_{j,t} = P_t^{\text{load}},\quad j \neq j^{\text{hdp}}
    \label{eq:load_bus_active}
\end{equation}
\begin{equation}
    - S_{\max}^{\text{sub}} \le P_{ij,t}, Q_{ij,t} \le S_{\max}
    \label{eq:line_rating_box}
\end{equation}
\begin{equation}
    P_{ij,t} - \sqrt{2} S_{\max} \le Q_{ij,t} \le P_{ij,t} + \sqrt{2} S_{\max}
    \label{eq:line_rating_upper}
\end{equation}
\begin{equation}
    - P_{ij,t} - \sqrt{2} S_{\max} \le Q_{ij,t} \le - P_{ij,t} + \sqrt{2} S_{\max}
    \label{eq:line_rating_lower}
\end{equation}
\begin{equation}
    V_{i,t}^{\text{sq}} - V_{j,t}^{\text{sq}}
    = 2 \left( r_{ij} P_{ij,t} + x_{ij} Q_{ij,t} \right)
    \label{eq:voltage_drop}
\end{equation}
\begin{equation}
    V_{\min}^{\text{sq}} \le V_{i,t}^{\text{sq}} \le V_{\max}^{\text{sq}}
    \label{eq:voltage_limit}
\end{equation}
\begin{equation}
    V_{i,t}^{\text{sq}} = V_{\text{sub}}^{\text{sq}}, \quad i = j^{\text{sub}}
    \label{eq:substation_voltage}
\end{equation}
where $i$, $j$, and $k$ are node indices; $j^{\text{sub}}$ and $j^{\text{hdp}}$ are the substation and HDP nodes; and $ij$ denotes the line connecting nodes $i$ and $j$. $\Omega_{j}^{\text{from}}$ and $\Omega_{j}^{\text{to}}$ are the sets of from- and to-nodes of node $j$. $P_{j,t}$ and $Q_{j,t}$ are the node active/reactive loads, and $P_{t}^{\text{hdp}}$ is the HDP net power consumption. $P_{ij,t}$ and $Q_{ij,t}$ represent the line active/reactive flows. $r_{ij}$ and $x_{ij}$ are the line resistance and reactance. $V_{i,t}^{\text{sq}}$ is the squared voltage magnitude at node $i$, bounded by $V_{\min}^{\text{sq}}$ and $V_{\max}^{\text{sq}}$, and $V_{\text{sub}}^{\text{sq}}$ is the substation squared voltage. $S_{\max}$ denotes the line apparent-power limit.

\section{Problem Formulation of Coordinated HDP-Power System Scheduling}

\subsection{Original Problem Formulation}
The objective of the coordinated HDP and distribution system is to minimize the total operating cost under market electricity prices while meeting water-demand and water-quality requirements and satisfying all operational constraints. The HDP net load can be negative when PV output is high; thus, power export is allowed and a selling price lower than the buying price is assumed to reflect a typical tariff structure in high renewable-penetration power systems, like the net billing tariff ``NEM~3.0" in California~\cite{nem3-lbl-2024}. The objective is formulated in \eqref{eq:dp_objective}, and \eqref{eq:hdp_power_split}--\eqref{eq:dp_power_nonneg} constrain buying and selling power. In summary, the original problem formulation for coordinated scheduling of the HDP and the power system is given by \eqref{eq:hpp_head}--\eqref{eq:dp_power_nonneg}, denoted as $\boldsymbol{\mathcal{P}_{\text{ORI}}}$.
\begin{equation}
    f^{\text{hdp}}=\min\nolimits\sum\nolimits_{t=1}^{T}
    \big(\rho_t^{+}P_t^{\text{hdp}+}-\rho_t^{-}P_t^{\text{hdp}-}\big)\Delta t
    \label{eq:dp_objective}
\end{equation}
\begin{equation}
    P_t^{\text{hdp}}
    = P_t^{\text{hdp}+} - P_t^{\text{hdp}-}
    \label{eq:hdp_power_split}
\end{equation}
\begin{equation}
    P_t^{\text{hdp}+}, P_t^{\text{hdp}-} \ge 0
    \label{eq:dp_power_nonneg}
\end{equation}
where $\rho_t^{+}$ and $\rho_t^{-}$ denote the electricity buying and selling prices at time $t$. $P_t^{\text{hdp}+}$ and $P_t^{\text{hdp}-}$ represent the HDP power absorbed from the grid and the power injected into the grid, respectively.

\vspace{-2mm}
\subsection{Problem Simplification}
Due to the on-off decision variable $U_t^{\text{on}}$, the pump functions \eqref{eq:hpp_head}--\eqref{eq:hpp_power}, the salt-separation constraints \eqref{eq:salt_balance}--\eqref{eq:ro_tds_ave}, and the tank-mixing equation \eqref{eq:tank_salt_balance}, the original problem $\boldsymbol{\mathcal{P}_{\text{ORI}}}$ is highly nonlinear and nonconvex, making it very challenging to solve. We note, however, that the permeate TDS $S_t^{\text{pe}}$ ($<1~\mathrm{kg/m^3}$) is much lower than the seawater TDS $S^{\text{fd}}$ ($30$--$45~\mathrm{kg/m^3}$). Therefore, the $S_t^{\text{pe}}$ terms on the right-hand sides of \eqref{eq:salt_balance} and \eqref{eq:perm_tds} are neglected, and \eqref{eq:osmotic_perm} is removed, which affects \eqref{eq:ro_osmotic_diff} and leads to the updated constraints \eqref{eq:salt_balance_simp}--\eqref{eq:osmotic_pressure_simp}. Correspondingly, \eqref{eq:ro_tds_ave} is replaced by the bivariate function in \eqref{eq:ro_tds_ave_simp}, which can be readily linearized. This simplification implicitly assumes a slightly higher concentrate-side TDS; thus, the scheduling solution requires a marginally higher feed pressure than necessary, yielding marginally higher freshwater production and lower permeate TDS (better water quality), which is favorable for operation.

\vspace{-12pt}
\begin{equation}
    S^{\text{fd}} F_t^{\text{fd}} = S_t^{\text{br}} F_t^{\text{br}}
    \label{eq:salt_balance_simp}
\end{equation}
\begin{equation}
    k_S^{\text{ro}} S_t^{\text{ro}} = S_t^{\text{pe}} F_t^{\text{pe}}
    \label{eq:perm_tds_simp}
\end{equation}
\begin{equation}
    \Delta \pi_t^{\text{ro}} =
    {k^{\text{os}}\!\left( S^{\text{fd}} + S_t^{\text{br}} \right)}/{2}
    \label{eq:osmotic_pressure_simp}
\end{equation}
\begin{equation}
    S_t^{\text{ro}}
    = \frac{2 S^{\text{fd}} F_t^{\text{fd}}}{F_t^{\text{fd}} + F_t^{\text{br}}}
    \label{eq:ro_tds_ave_simp}
\end{equation}

Since we are concerned with the salt mass in the permeate flow, i.e., $S_t^{\text{pe}} F_t^{\text{pe}}$, we introduce a new variable $M_t^{\text{Spe}}$ to represent this term. Similarly, the variable $M_t^{\text{Stk}}$ is introduced. Accordingly, constraints~\eqref{eq:perm_tds_simp} and~\eqref{eq:tank_salt_balance} are reformulated as~\eqref{eq:salt_mass_perm} and~\eqref{eq:tank_salt_balance_mod}, respectively. It should be noted that the variable $S_t^{\text{pe}}$, which represents permeate water quality, is removed. Instead, two constraints are introduced to explicitly regulate water quality by controlling the salt mass in the permeate flow: constraint~\eqref{eq:perm_salt_limit} ensures that the permeate TDS remains within the required range, while constraint~\eqref{eq:perm_salt_off} enforces that no salt is transported when the desalination system is turned off. We then obtained a simplified problem formulation for the coordinative scheduling of the HDP and the power system, $\boldsymbol{\mathcal{P}_{\text{SP}}}$. 

\vspace{-6pt}
\begin{equation}
    k_S^{\text{ro}} S_t^{\text{ro}} = M_t^{\text{Spe}}
    \label{eq:salt_mass_perm}
\end{equation}
\begin{equation}
    M_t^{\text{Stk}}
    = M_{t-1}^{\text{Stk}}
    + M_t^{\text{Spe}} \Delta t
    - S_t^{\text{out}} F_t^{\text{out}} \Delta t
    - S_{\text{ave}}^{\text{tk}} W_t^{\text{flush}}
    \label{eq:tank_salt_balance_mod}
\end{equation}
\begin{equation}
    0 \le M_t^{\text{Spe}} \le S_{\max}^{\text{pe}} F_t^{\text{pe}}
    \label{eq:perm_salt_limit}
\end{equation}
\begin{equation}
    0 \le M_t^{\text{Spe}} \le S_{\max}^{\text{pe}} F_{\max}^{\text{pe}} U_t^{\text{on}}
    \label{eq:perm_salt_off}
\end{equation}
\begin{equation}
    M_t^{\text{Stk}} = S_t^{\text{tk}} W_t^{\text{tk}}
    \label{eq:salt_mass_tk}
    \vspace{-6pt}
\end{equation}

\vspace{-2mm}
\subsection{Linearization}
The remaining nonlinear constraints include the pump operation functions~\eqref{eq:hpp_head}--\eqref{eq:hpp_power}, the salt-separation-related constraints~\eqref{eq:salt_balance_simp} and~\eqref{eq:ro_tds_ave_simp}, and the tank-mixing constraint~\eqref{eq:salt_mass_tk}. These bivariate constraints can be linearized using one-dimensional, rectangular, and triangular piecewise linearization~\cite{math-2010-linear-tri}. Although~\cite{math-2010-linear-tri} recommends the rectangular method, our tests show that triangular linearization performs best. This is because several nonlinear constraints share common variables, such as $F_t^{\text{fd}}$ and $\omega_t$ on the left-hand sides of~\eqref{eq:hpp_head}--\eqref{eq:hpp_power} and $F_t^{\text{br}}$ in~\eqref{eq:salt_balance_simp} and~\eqref{eq:ro_tds_ave_simp}, keeping the triangular-linearized problem size modest. Moreover, triangular linearization avoids the segment misalignment that can occur with rectangular linearization, which can cause infeasibility in some cases.

Triangular linearization is a convex-combination method. Consider a bivariate function $z = f(x,y)$. Let there be $M$ and $N$ breakpoints (including extreme points) along the $x$- and $y$-dimensions, indexed by $m$ and $n$, respectively. This results in $M-1$ and $N-1$ segments along each dimension and $(M-1)(N-1)$ piecewise-linear surface patches in total. Let the continuous variable $\lambda_{m,n,t}$ denote the convex-combination coefficient associated with a vertex of a surface patch, where $\lambda_{m,n,t} \in [0,1]$. Because the four vertices of each surface patch generally do not lie on a flat plane, each patch is divided into an upper triangle and a lower triangle. The selection of these triangles is indicated by the binary variables $u_{m,n,t}^{\text{up}}$ and $u_{m,n,t}^{\text{lo}}$. The activation of vertices is bounded by the desalination operating status through \eqref{eq:tri_sum_lambda} and \eqref{eq:tri_lambda_bound}. The values of $x$, $y$, and $z$ are determined by \eqref{eq:tri_x}--\eqref{eq:tri_z}. When the desalination process is on, the operating point is restricted to lie on only one triangle, as enforced by \eqref{eq:tri_one_triangle}. Constraint \eqref{eq:tri_sos3} ensures that a nonzero $\lambda_{m,n,t}$ is associated only with the vertices of the selected triangle, thereby enforcing $\lambda_{m,n,t}$ to be a special ordered set of type 3 (SOS3). Finally, the outer boundaries of the feasible region are forced to zero using \eqref{eq:tri_boundary_n} and \eqref{eq:tri_boundary_m}. With the linearization, the problem is reformulated as a tractable MILP problem, $\boldsymbol{\mathcal{P}_{\text{SPLN}}}$.

\vspace{-2mm}
\begin{equation}
    \sum\nolimits_{m=1}^{M} \sum\nolimits_{n=1}^{N} \lambda_{m,n,t} = U_t^{\text{on}}
    \label{eq:tri_sum_lambda}
\end{equation}
\begin{equation}
    0 \le \lambda_{m,n,t} \le U_t^{\text{on}}
    \label{eq:tri_lambda_bound}
\end{equation}
\begin{equation}
    x_t = \sum\nolimits_{m=1}^{M} \sum\nolimits_{n=1}^{N}
    \lambda_{m,n,t} \, x_m
    \label{eq:tri_x}
\end{equation}
\begin{equation}
    y_t = \sum\nolimits_{m=1}^{M} \sum\nolimits_{n=1}^{N}
    \lambda_{m,n,t} \, y_n
    \label{eq:tri_y}
\end{equation}
\begin{equation}
    z_t = \sum\nolimits_{m=1}^{M} \sum\nolimits_{n=1}^{N}
    \lambda_{m,n,t} \, z_{m,n}
    \label{eq:tri_z}
\end{equation}
\begin{equation}
    \sum\nolimits_{m=1}^{M-1} \sum\nolimits_{n=1}^{N-1}
    \left( u_{m,n,t}^{\text{up}} + u_{m,n,t}^{\text{lo}} \right)
    = U_t^{\text{on}}
    \label{eq:tri_one_triangle}
\end{equation}
\begin{equation}
    \begin{split}
    \lambda_{m+1,n+1,t}
    \le\;&
    u_{m+1,n+1,t}^{\text{up}}
    + u_{m+1,n,t}^{\text{up}}
    + u_{m,n,t}^{\text{up}} \\
    &+
    u_{m+1,n+1,t}^{\text{lo}}
    + u_{m,n+1,t}^{\text{lo}}
    + u_{m,n,t}^{\text{lo}}
    \end{split}
\label{eq:tri_sos3}
\end{equation}
\begin{equation}
    u_{m,N,t}^{\text{up}} = 0, \quad
    u_{m,N,t}^{\text{lo}} = 0,
    \quad m = 1,\ldots,M
    \label{eq:tri_boundary_n}
\end{equation}
\begin{equation}
    u_{M,n,t}^{\text{up}} = 0, \quad
    u_{M,n,t}^{\text{lo}} = 0,
    \quad n = 1,\ldots,N
    \label{eq:tri_boundary_m}
\end{equation}

\begin{algorithm}[!h]
    \small
    \caption{TDCSO scheduling: $\boldsymbol{\mathcal{P}^{\text{SO}}_{\text{SPLN}}}$}
    \label{alg:tdcso}
    \begin{algorithmic}[1]
    
    \STAGE{Step 1: Commitment scheduling $\boldsymbol{\mathcal{P}}^{\mathrm{SO\text{-}1}}_{\mathrm{SPLN}}\!\left(\boldsymbol{x}_t^{\mathrm{w}},\,\boldsymbol{x}_{t,s}^{\mathrm{p}},\,U_t^{\mathrm{on}}\right)$}
    \STATE \hspace{1.5em}\parbox[t]{0.88\linewidth}{Build the operation model by modifying $\boldsymbol{\mathcal{P}_{\text{SPLN}}}$: embed $\Omega_s$ into PV constraints~\eqref{eq:pv_power_limits}--\eqref{eq:pv_inverter_limit} and power system constraints~\cref{eq:node_power_balance,eq:hdp_power_injection,eq:hdp_power_coupling,eq:load_bus_active,eq:line_rating_box,eq:line_rating_upper,eq:line_rating_lower,eq:voltage_drop,eq:voltage_limit,eq:substation_voltage} and their associated variables, disable mixing flexibility by $S^{\text{pe}}_{\max}=S^{\text{out}}_{\max}$.}

    \STATE \hspace{1.5em}\parbox[t]{0.88\linewidth}{Solve the problem and obtain the desalination commitment $\tilde{U}_t^{\text{on}}$ and the flushing scheme.}

    \vspace{2mm}
    \STAGE{Step 2: Stochastic flexibility scheduling with fixed commitment $\boldsymbol{\mathcal{P}}^{\mathrm{SO\text{-}2}}_{\mathrm{SPLN}}\!\left(\boldsymbol{x}_{t,s}^{\mathrm{w}},\,\boldsymbol{x}_{t,s}^{\mathrm{p}}\,\middle|\,U_t^{\mathrm{on}}=\tilde{U}_t^{\mathrm{on}}\right)$}
    \STATE \hspace{1.5em}\parbox[t]{0.88\linewidth}{For each scenario $s$, build the operation model by modifying $\boldsymbol{\mathcal{P}_{\text{SPLN}}}$: exclude flush related~\cref{eq:flush_water,eq:flush_energy,eq:shutdown_logic_1,eq:shutdown_logic_1,eq:shutdown_logic_2,eq:start_logic_1,eq:start_logic_2,eq:shutdown_min}, enable mixing flexibility by allowing $S^{\text{pe}}_{\max}>S^{\text{out}}_{\max}$, use the obtained commitment $U_t^{\text{on}}=\tilde{U}_t^{\text{on}}$.}
    \STATE \hspace{1.5em}\parbox[t]{0.88\linewidth}{Solve the operation model of each scenario $s$ in parallel, and obtain the expected cost:
    $\min \sum\nolimits_{s\in\Omega_s} o_s \sum\nolimits_{t=1}^{T}\big(\rho_{t,s}^{+}P_{t,s}^{\text{hdp}+}-\rho_{t,s}^{-}P_{t,s}^{\text{hdp}-}\big)\Delta t$}
    \end{algorithmic}
\end{algorithm}

\vspace{-2mm}
\subsection{Two-step Stochastic Scheduling}
This work accounts for uncertainties in intermittent PV generation and time-varying electricity prices. However, fully exploiting the unlocked flexibility in $\boldsymbol{\mathcal{P}_{\text{SPLN}}}$ already increases the computational burden, and explicitly embedding these uncertainties further exacerbates it. To incorporate uncertainty while maintaining tractability, we propose a TDCSO stochastic scheduling algorithm, $\boldsymbol{\mathcal{P}^{\text{SO}}_{\text{SPLN}}}$, which decomposes the scheduling task into (i) commitment determination (on-off status) and (ii) stochastic flexibility scheduling with fixed commitment, as shown in Algorithm~\ref{alg:tdcso}, where $\boldsymbol{x}_{t,s}^{\text{p}}$ denotes the variables associated with PV operation and power-system operation in scenario $s$, while $\boldsymbol{x}_{t,s}^{\text{w}}$ denotes the desalination-related variables except the commitment variable $U_{t}^{\text{on}}$. To construct the uncertainty set, we first generate $N_s^{\text{ori}}$ scenarios using a Copula-based joint scenario generation method based on PV and energy-price forecasts and their deviation ranges, and then apply $k$-medoids clustering to obtain the reduced scenario set $\Omega_s$ and the corresponding scenario probabilities $o_s$. The Step 1 problem $\boldsymbol{\mathcal{P}}^{\mathrm{SO\text{-}1}}_{\mathrm{SPLN}}$ enforces the same desalination decisions across all scenarios, while incorporating uncertainty on the PV and power-system sides. It is solved with mixing flexibility disabled to determine the desalination on-off status. Then, the Step 2 problem, $\boldsymbol{\mathcal{P}}^{\mathrm{SO\text{-}2}}_{\mathrm{SPLN}}$, reschedules the system with mixing flexibility enabled by fixing the obtained on-off status and incorporating uncertainties into all operational decisions, it can be solved effectively in parallel.

\section{Simulation Results}
\begin{figure}[!ht]
    \vspace{-10pt}
    \centering
    \includegraphics[width=1.0\linewidth]{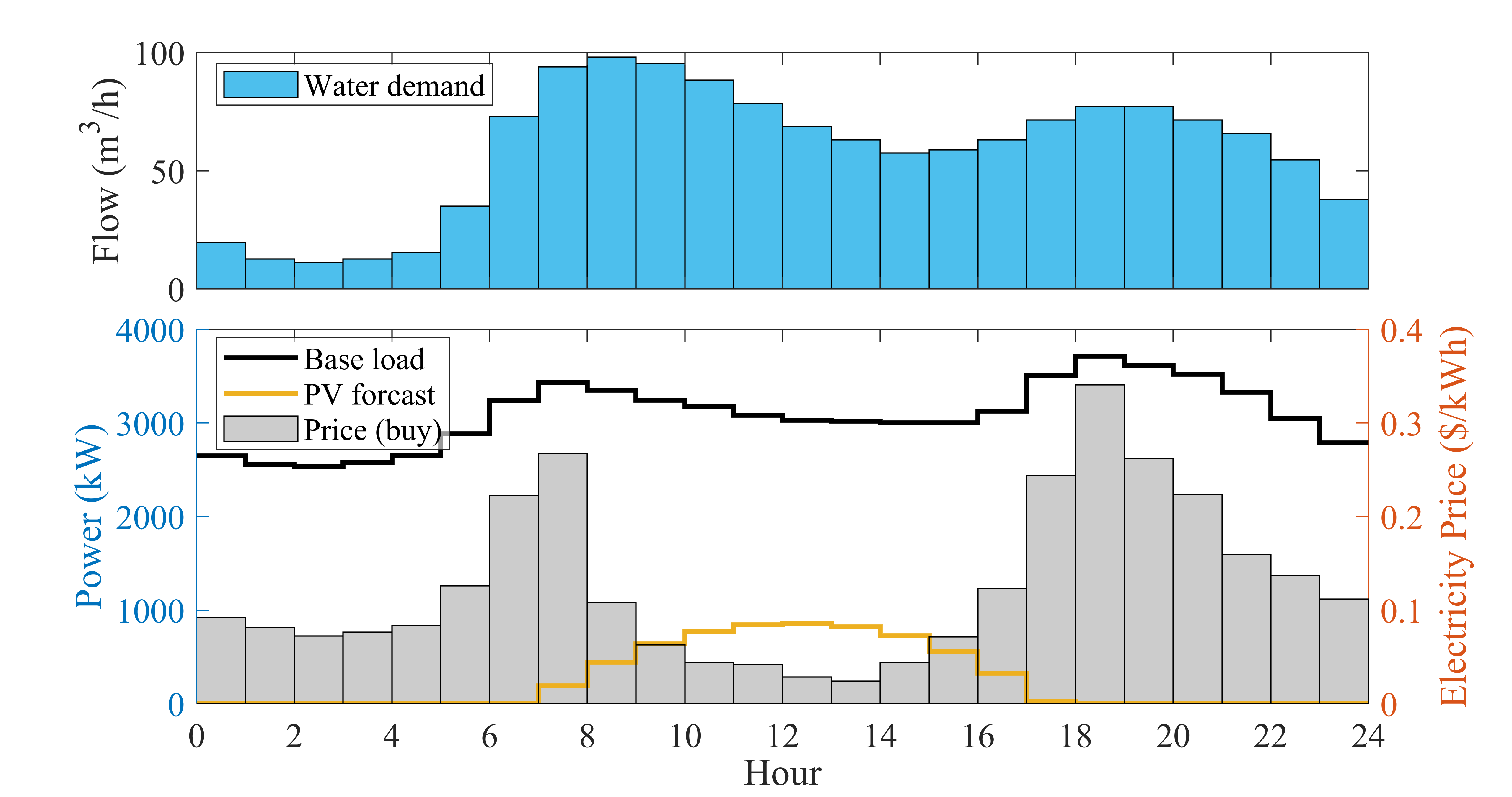}
    \vspace{-4mm}
    \caption{Water demand (top), forecast of the co-located PV, the load of the distribution system except the HDP, and the buy price of energy (bottom).}
    \label{fig:case_inputs}
\end{figure}

The test system, as shown in Fig.~\ref{fig:sys_layout}, uses the 33-bus test feeder~\cite{math-2002-distflow-33}, with an HDP co-located with a 1000~kW PV system. The RO operating pressure ranges from 6000 to 6500~kPa. The freshwater storage tank capacity is 1800~m$^{3}$, with a state of charge (SOC) range of 0.2--1.0 and an initial SOC of 0.4. The seawater TDS is 42~kg/m$^{3}$. To meet end-user water quality requirements, the outflow TDS must not exceed 0.35~kg/m$^{3}$~\cite{usbr-2014-vary-tds}, while the initial tank TDS is 0.3~kg/m$^{3}$. Daily water demand is 1400~m$^{3}$ with the hourly profile in Fig.~\ref{fig:case_inputs}~\cite{utah-2019-optimal-desali}. The HPP is rated at 600~kW and operates with a normalized speed range of 0.7--1.3 (VFD-enabled), motor and VFD efficiencies are 0.95 and 0.97, respectively, and the maximum pump flow rate at nominal speed is 250~m$^{3}$/hr. Electricity buying price~\cite{utah-2019-optimal-desali}, base load excluding the HDP, and PV forecast~\cite{pv_load_2024_my} are shown in Fig.~\ref{fig:case_inputs}, the selling price is assumed to be half of the buying price. All system parameters and case-study data are available in~\cite{data-upload-github}. The scheduling problem is solved with Gurobi~12 in MATLAB~R2023 on an Intel Core i9-14900K CPU desktop with 128~GB RAM, using the default MIP gap of 0.01\%.

\begin{figure}[!ht]
    \vspace{-5 pt}
    \centering
    \includegraphics[width=1.0\linewidth]{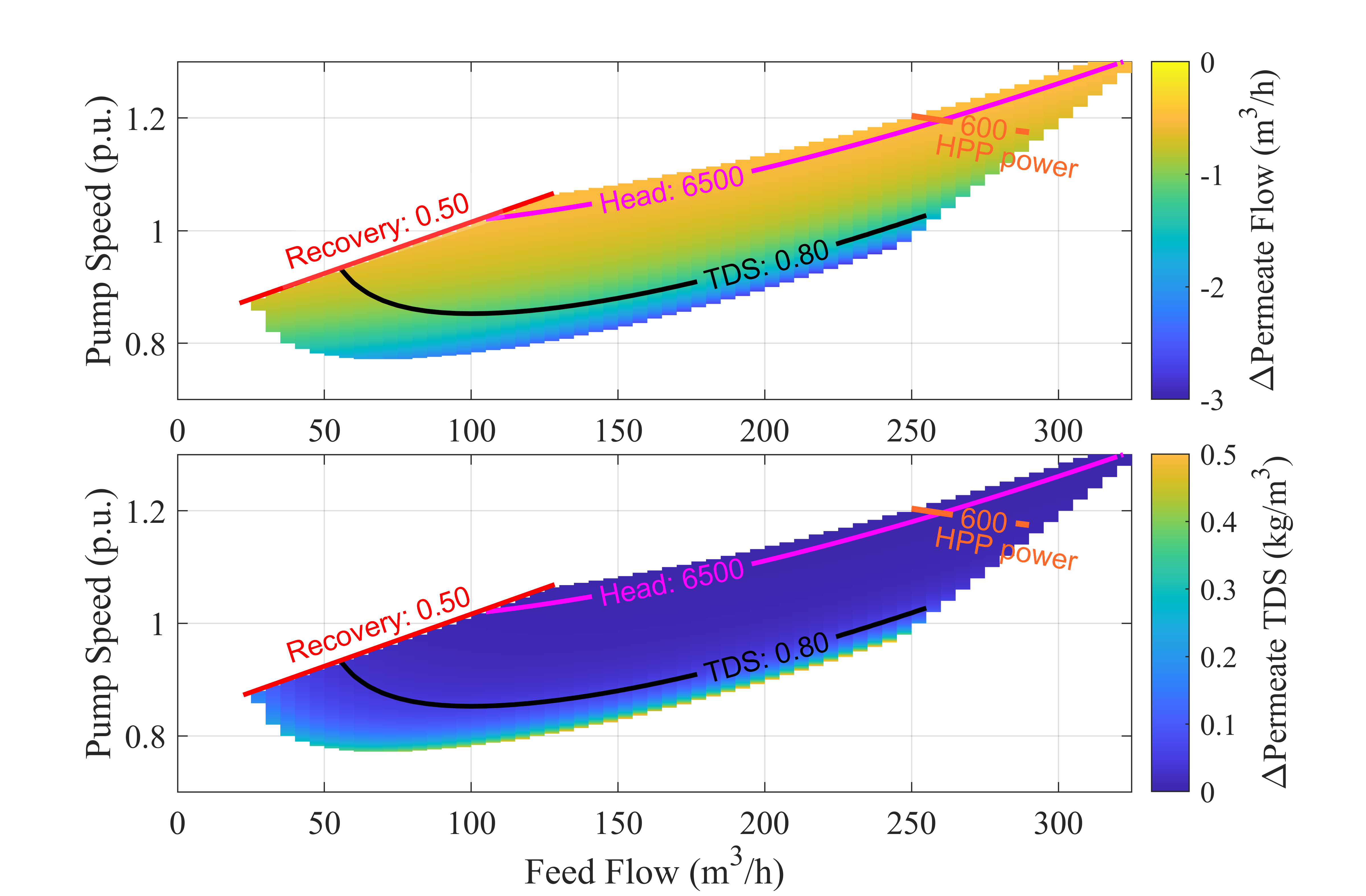}
    \caption{Calculation errors of the simplified model for permeate flow (top) and TDS (bottom), computed as (simplified -- full).}
    \label{fig:case_model_error}
    \vspace{-2mm}
\end{figure}

\vspace{-3mm}
\subsection{Simplification Evaluation}
\label{subsec:simp_evalu}
By simulating multiple operating points and directly solving the non-linearized problems using BARON, the simplified desalination model $\boldsymbol{\mathcal{P}_{\text{SP}}}$ and the full desalination model $\boldsymbol{\mathcal{P}_{\text{ORI}}}$ are compared to analyze model errors in two key outputs: permeate TDS and permeate water flow, as shown in Fig.~\ref{fig:case_model_error}. Note that the tank storage and power system components are excluded in this comparison, and the operational constraints in the simulations are slightly relaxed compared with those used in the subsequent case studies. The feasible operating region is illustrated by the circled central area, which is bounded by the operational limits on the maximum recovery rate (left), feed head (top), HPP power (top right), feed flow (right), and permeate TDS (bottom). Within this flexibility region, the model errors are very small, remaining within $-1~\text{m}^3/\text{hr}$ for permeate water flow and $0.05~\text{kg}/\text{m}^3$ for permeate TDS. Larger errors occur primarily along the lower boundary of the feasible region. If the maximum allowable permeate TDS is raised to $2~\text{kg}/\text{m}^3$, the errors can reach approximately $-2.5~\text{m}^3/\text{hr}$ for permeate water flow and $0.5~\text{kg}/\text{m}^3$ for TDS. This discrepancy arises because the simplified model assumes a higher concentrate-side TDS, yielding slightly higher predicted permeate TDS and lower permeate water production.

\vspace{-2mm}
\subsection{Performance with Unlocked Flexibility}
Four cases are compared based on model capabilities and operational flexibility strategies. \case{NoMix}: only water-storage flexibility is considered, with permeate TDS limited to $0.35~\text{kg/m}^3$. Tank mixing is not considered. This operation strategy is adopted in~\cite{kfupm-2022-risk} and serves as the base case. \case{MixIni}: same as \case{NoMix}, except that tank mixing is enabled and the tank end-TDS must not exceed the initial tank TDS. \case{MixFlex}: both water-storage and salinity flexibilities are considered, with permeate TDS allowed up to $0.80~\text{kg/m}^3$, enabling additional flexibility via proactive storage salinity management. Tank mixing is included, and there is no restriction on the tank end-TDS. \case{MixFlexIni}: same as \case{MixFlex}, except that the tank end-TDS must not exceed the initial tank TDS.

Table~\ref{tab:ems_cmp} shows the scheduling solutions obtained using the proposed $\boldsymbol{\mathcal{P}_{\text{SPLN}}}$ and their performance verified by executing schedules in the full desalination model $\boldsymbol{\mathcal{P}_{\text{ORI}}}$. We observe that:

\subsubsection{Tank TDS}
The \case{NoMix} model does not include tank mixing and therefore cannot capture tank-TDS mixing dynamics. As a result, it cannot ensure that the tank end-TDS does not exceed its initial value, leading to higher permeate TDS by the end of the day. If not later reduced by low TDS permeate water, tank water quality can gradually violate the requirement, which also implies a shadow cost because producing lower-TDS permeate flow requires higher specific energy consumption (SEC, in kWh/m$^{3}$ of permeate) and thus higher operating cost. \case{MixFlex} faces a similar limitation in terms of lacking end-TDS regulation. In contrast, \case{MixIni} and \case{MixFlexIni} explicitly model and regulate tank end-TDS, ensuring it does not exceed the initial value. Moreover, all three mixing included cases produce a lower tank end-TDS than predicted by the scheduling model, which is expected as the simplified formulation is used for scheduling, as analyzed in~\ref{subsec:simp_evalu}. Consequently, real operation produces slightly more water and lower permeate TDS, which is favorable.

\subsubsection{Freshwater}
The total water scheduled production  are the same across all four cases. They all yield slightly higher actual water production due to the simplified modeling of the desalination process. In addition, the scheduled water production in all cases is higher than the water demand because of flushing consumption, which can be clearly observed in the detailed scheduling results in the following Subsection~\ref{subsec:scheduling_results}.

\subsubsection{Cost}
Since actual water production exceeds the scheduled, we report a prorated cost, which represents the verified cost scaled to the daily demand of 1400~m$^3$. Compared with \case{NoMix}, \case{MixIni} incurs a higher cost because it regulates the tank end-TDS, the same holds for \case{MixFlex} versus \case{MixFlexIni}. However, it is difficult to isolate the incremental cost of guaranteeing the end-TDS, as the desalination scheduling is influenced by water demand, PV output, and electricity prices. A fair and meaningful comparison is between \case{MixIni} and \case{MixFlexIni}, since both enforce the end-TDS constraint and thus incur no shadow cost from increased tank end-TDS in real operation. \case{MixFlexIni} reduces the prorated cost by \$13.72 (2.36\%) relative to \case{MixIni}, demonstrating the benefit of flexibility enabled by proactive tank TDS management.

\subsubsection{Computation}
Integrating tank mixing and end-TDS regulation increases computational burden: 15.8~s for \case{NoMix}, 128.6~s for \case{MixIni}, 38.9~s for \case{MixFlex}, and 264.9~s for \case{MixFlexIni}. Introducing proactive TDS flexibility further increases runtime:~\case{MixFlex} and \case{MixFlexIni} require about 2.5 and 2.1 times the runtime of \case{NoMix} and \case{MixIni}, respectively, due to the larger feasible region under the higher permeate-TDS bound. However, all runtimes remain well within practical limits and do not pose any obstacle for day-ahead implementation.

\subsubsection{Optimality} 
We attempted to solve the original problem~$\boldsymbol{\mathcal{P}_{\text{ORI}}}$ using global solvers, but computation was prohibitively slow; for instance, BARON reached only 30\% progress after six hours. Therefore, the optimality of the simplified and linearized problem~$\boldsymbol{\mathcal{P}_{\text{SPLN}}}$ is verified by a MILP scheduling problem that picks operation points from preoccupied feasible operation points (FOPs). For the FOP method, the feed-flow and pump-speed domains are partitioned into segments with widths of 5~m$^3$/hr and 0.002, respectively. Under this segmentation, the maximum power approximation error is about 0.67\% when the normalized pump speed is 1. With these settings, the FOP can obtain near-optimal solutions. Table~\ref{tab:ems_opti_verify} shows the results obtained using the FOP method. Compared with the FOP results, the verified costs of the four cases using the proposed method are 1.01\%, 1.14\%, 0.61\%, and 1.32\% higher, respectively. For the prorated cost, the differences are -0.04\%, 0.07\%, -0.46\%, and 0.27\%. These results demonstrate that the optimality of the proposed~$\boldsymbol{\mathcal{P}_{\text{SPLN}}}$ is sufficiently good for practical implementation.

\begin{table}[!t]
\centering
\caption{Scheduling Solutions and Performance Validation}
\label{tab:ems_cmp}
\setlength{\tabcolsep}{4pt} 
\renewcommand{\arraystretch}{1.15}
    \begin{tabularx}{\linewidth}{l l c c c c}
    \toprule
    \textbf{Case} &  & \textbf{NoMix} & \textbf{MixIni} & \textbf{MixFlex} & \textbf{MixFlexIni} \\
    \midrule
    
    \multirow{2}{*}{\begin{tabular}[c]{@{}c@{}}
    Tank end-TDS \\
    (kg/m$^3$)
    \end{tabular}}
     & Scheduled      & --     & 0.3000 & \textcolor{red}{0.3113} & 0.3000 \\
     & Verified & \textcolor{red}{0.3031} & 0.2976 & \textcolor{red}{0.3053} & 0.2965 \\
    \midrule
    
    \multirow{2}{*}{\begin{tabular}[c]{@{}c@{}}
    Water production \\
    (m$^3$)
    \end{tabular}}
     & Scheduled      & 1430.00   & 1430.00   & 1430.00   & 1430.00 \\
     & Verified & 1444.96 & 1445.32 & 1445.45 & 1445.00 \\
    \midrule
    
    \multirow{3}{*}{Cost (\$)}
     & Scheduled      & 579.9  & 590.32 & 569.75 & 578.41 \\
     & Verified & 578.6  & 588.69 & 565.66 & 574.70 \\
     & Prorated & 572.6  & 582.45 & 559.62 & 568.73 \\
    \midrule
    
    Calcu time (s)
     & Scheduled & 15.8 & 128.6 & 38.9 & 264.9 \\
    \bottomrule
    \end{tabularx}
\end{table}

\begin{table}[!t]
\caption{Optimality Validation of Scheduling Solutions by FOP-based Method}
\label{tab:ems_opti_verify}
\centering
\setlength{\tabcolsep}{4pt}
\renewcommand{\arraystretch}{1.2}
    \begin{tabular}{l c c c c}
    \toprule
    \textbf{Case} & \textbf{NoMix} & \textbf{MixIni} & \textbf{MixFlex} & \textbf{MixFlexIni} \\
    \hline
    Tank end-TDS (kg/m$^3$)        & \textcolor{red}{0.3066} & 0.3001 & \textcolor{red}{0.3061} & 0.3002 \\
    Water production (m$^3$)  & 1430.11 & 1430.10 & 1430.40 & 1430.10 \\
    Cost (\$)                 & 572.84 & 582.06 & 562.21 & 567.19 \\
    \bottomrule
    \end{tabular}
\end{table}

\begin{figure}[!t]
    \centering
    \includegraphics[width=1.0\linewidth]{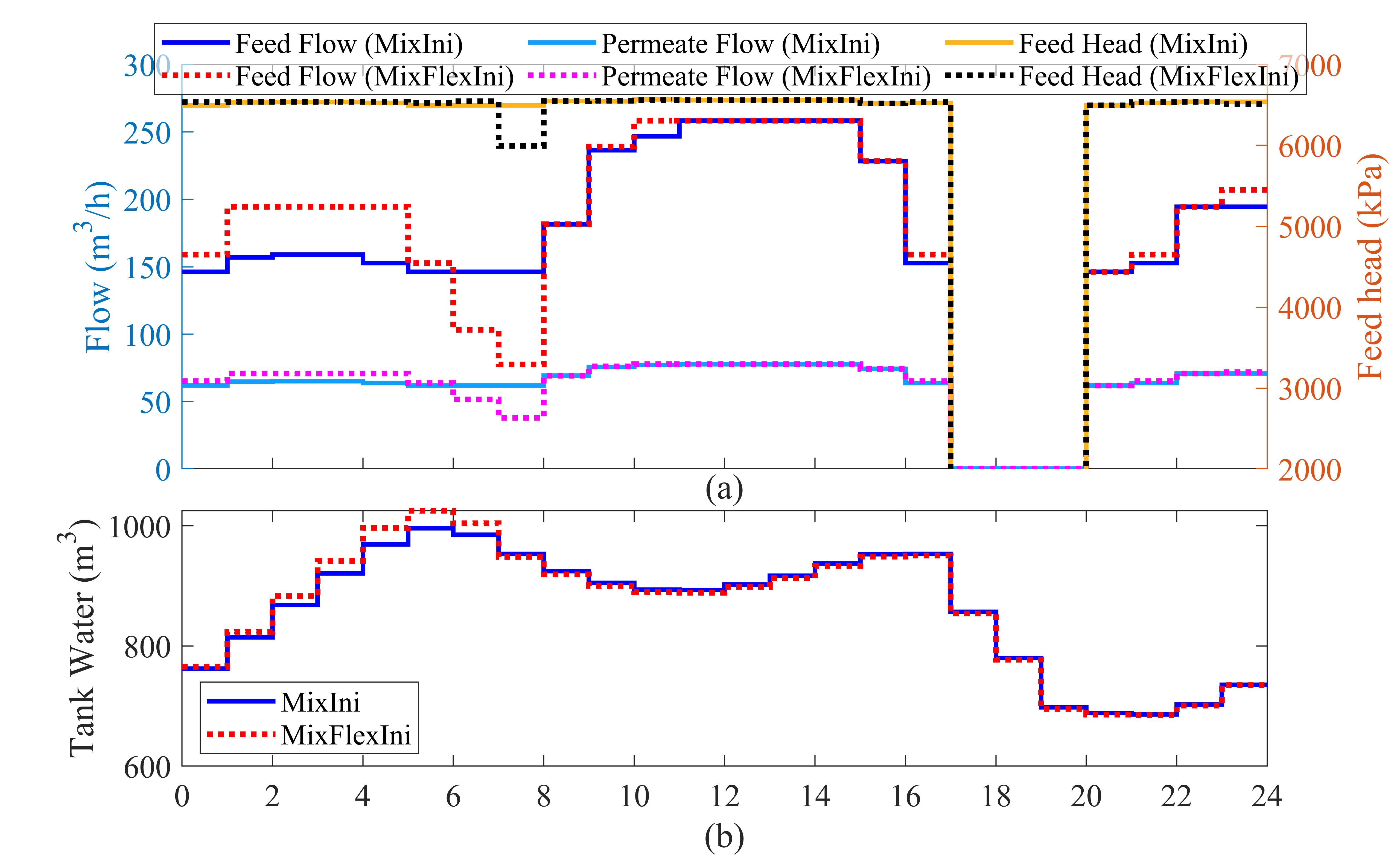}
    \vspace{-6mm}
    \caption{(a) Feed flow and feed pressure, and (b) tank water storage.}
    \label{fig:cmp_flow_head}
\end{figure}

\begin{figure}[!t]
    \centering
    \includegraphics[width=1.0\linewidth]{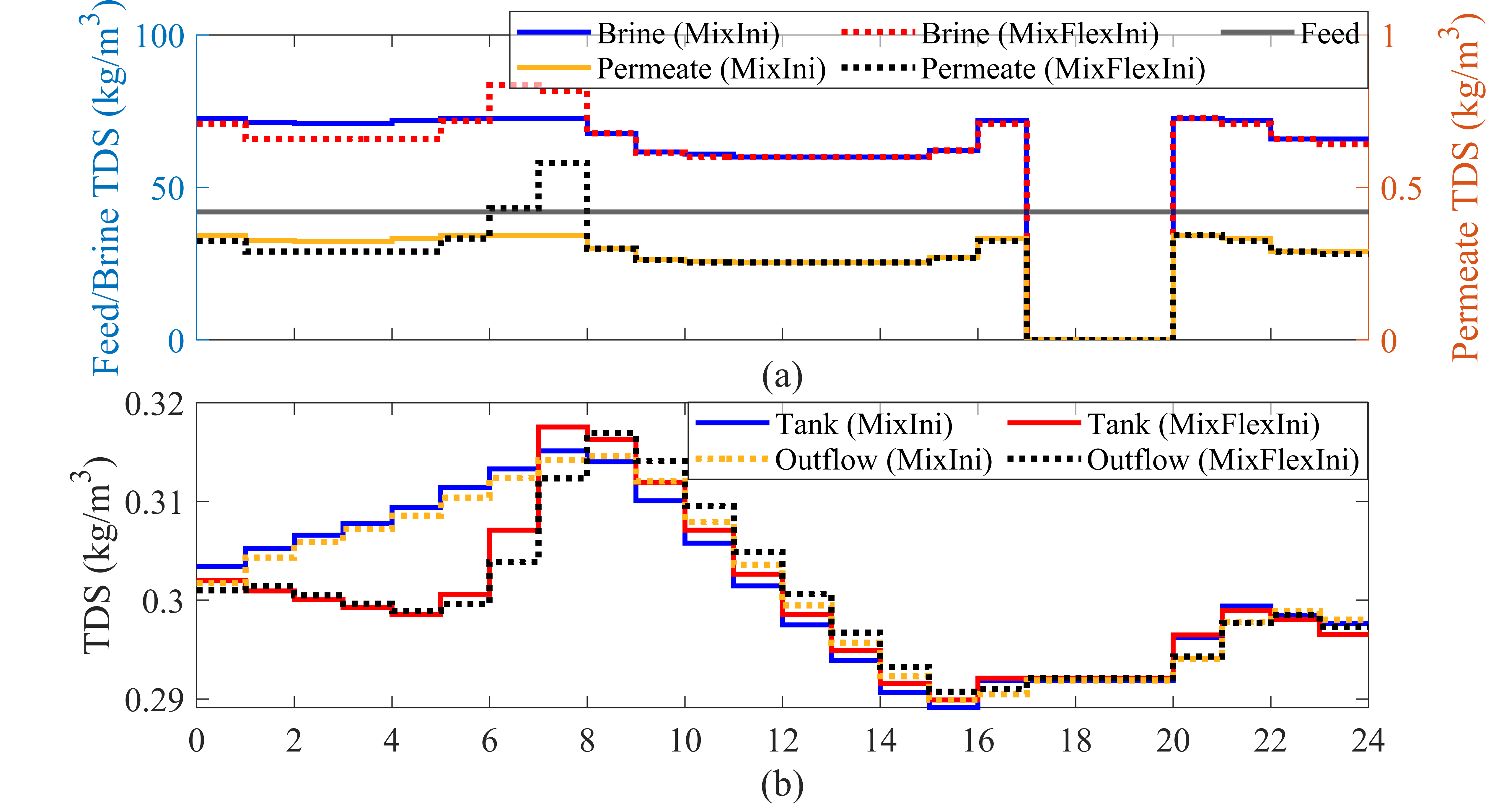}
    \vspace{-6mm}
    \caption{TDSs: (a) brine and permeate, and (b) tank water storage and outflow.}
    \label{fig:cmp_tds}
    \vspace{-6pt}
\end{figure}

\begin{figure}[!t]
    \centering
    \includegraphics[width=1.0\linewidth]{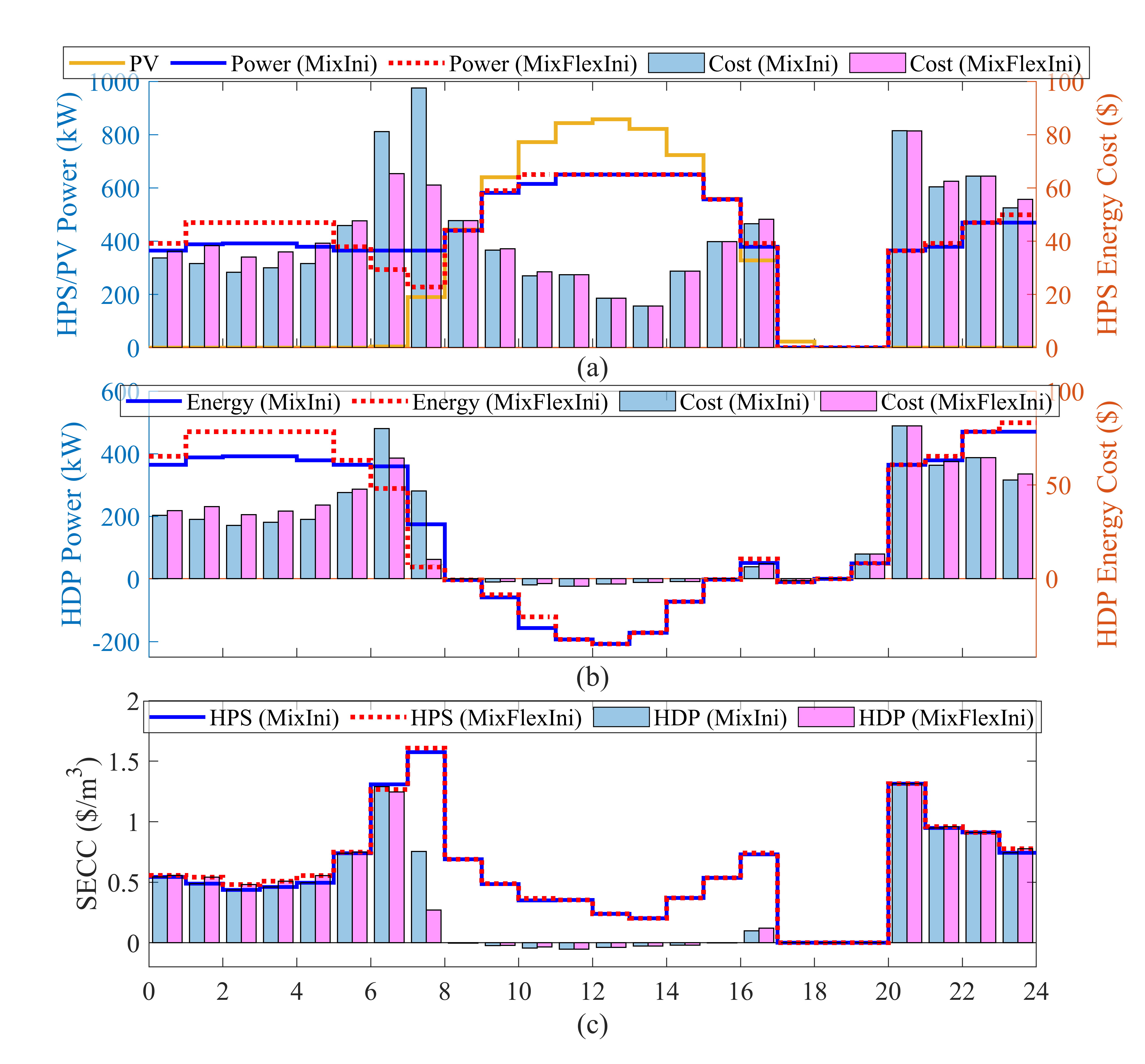}
    \vspace{-6mm}
    \caption{(a) HPS energy consumption and cost, (b) HDP net energy consumption and cost with flushing included, and (c) SECC of HPS and HDP.}
    \label{fig:cmp_hps_energy}
\end{figure}

\begin{figure}[!t]
    \vspace{-4pt}
    \centering
    \includegraphics[width=1.0\linewidth]{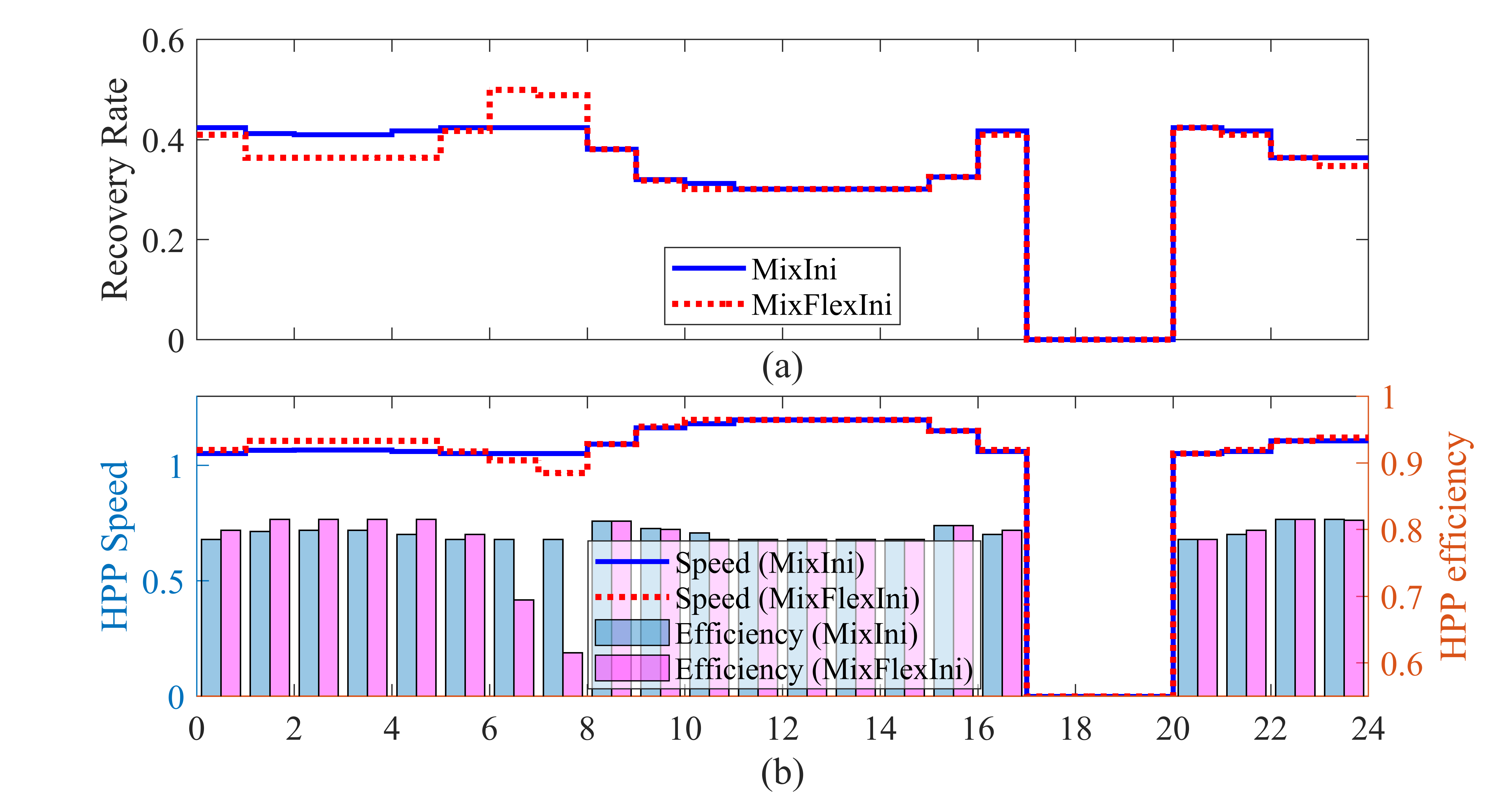}
    \vspace{-6mm}
    \caption{(a) Recovery rate, and (b) speed and efficiency of the HPP.}
    \label{fig:cmp_rec_eff}
    \vspace{-4pt}
\end{figure}

\subsection{Comparisons of the Scheduling}
\label{subsec:scheduling_results}
In this section, we compare the detailed verified performance of the scheduling solutions for \case{MixIni} and \case{MixFlexIni}, and further elaborate on the advantages of the proposed method.

As shown in Fig.~\ref{fig:cmp_flow_head}(a), in both cases the pressure varies only slightly, whereas the feed flow changes substantially, and desalination is shut down during the high-price hours 18--20. Compared with \case{MixIni}, \case{MixFlexIni} pumps less feed flow and operates at lower pressure around hour 8, coinciding with the second price peak, due to proactive flexibility management. It also produces more water before hour 7 to offset reduced production later (see Fig.~\ref{fig:cmp_flow_head}(b)), taking advantage of low electricity prices in the early hours.

Fig.~\ref{fig:cmp_tds} depicts the TDS of flows and water storage. The TDS of the brine flow and permeate flow is inversely correlated with the corresponding flow rate, as the pressures remain relatively stable. In addition, \case{MixFlexIni} produces lower TDS permeate water than \case{MixIni} during hours 1--5, while generating higher TDS permeate during the high-price period in hours 7--8. The resulting difference in tank water storage can be seen in Fig.~\ref{fig:cmp_tds}(b), where the tank TDS of \case{MixFlexIni} decreases during hours 1--5 and then increases during hours 6--8. After that, only a slight TDS difference remains between \case{MixFlexIni} and \case{MixIni}. Furthermore, the outflow TDS slightly deviates from the tank TDS, while the user-end water quality is well ensured; in particular, \case{MixFlexIni} provides better water quality to users during hours 1--8. This demonstrates the flexibility of TDS regulation in the closed loop of desalination and water storage.

Net energy consumption and the cost of the HPS and the HDP are shown in Fig.~\ref{fig:cmp_hps_energy}. For \case{MixFlexIni}, the HPS consumes less energy than \case{MixIni} during hours 7--8, which leads to a significant cost reduction due to the high energy price, as shown in Fig.~\ref{fig:cmp_hps_energy}(a). However, this operation comes at the expense of lower HPP efficiency. As shown in Fig.~\ref{fig:cmp_rec_eff}(a), despite achieving the highest recovery rate, reduced efficiency results in a slightly higher HPS specific energy consumption cost (SECC) for \case{MixFlexIni} (see Fig.~\ref{fig:cmp_hps_energy}(c)). Moreover, in hour 8, the reduction in net energy consumption and cost of the HDP is much more significant than that of the HPS, due to the considerable PV output. Furthermore, the energy consumption and the associated cost required for flushing in hours 18 and 20 can be clearly observed in Fig.~\ref{fig:cmp_hps_energy}(b), and they are not negligible.

Both \case{MixIni} and \case{MixFlexIni} take advantage of the PV power when it is available, since it has no cost and its selling price is half of the energy buying price. Driven by the availability of PV and the lowest electricity prices, the HPS power consumption is highest during hours 10--15, although it is accompanied by reduced HPP efficiency and recovery rate (see Fig.~\ref{fig:cmp_rec_eff}(b)). The benefit, however, is that the permeate flow reaches its highest level and the permeate TDS is the lowest, as shown in Figs.~\ref{fig:cmp_flow_head} and~\ref{fig:cmp_tds}. In addition, it is interesting to note that \case{MixFlexIni} achieves more PV self-consumption in hour 11 and thus reduces the power injection to the grid (see Fig.~\ref{fig:cmp_hps_energy}(b)). The reason is that it needs low-TDS permeate water to reduce the tank TDS (see Fig.~\ref{fig:cmp_tds}(b)), which is higher than that of \case{MixIni}, in order to ensure that the tank end-TDS does not exceed the initial tank TDS.

\begin{figure}[!ht]
    \vspace{-6pt}
    \centering
    \includegraphics[width=0.8\linewidth]{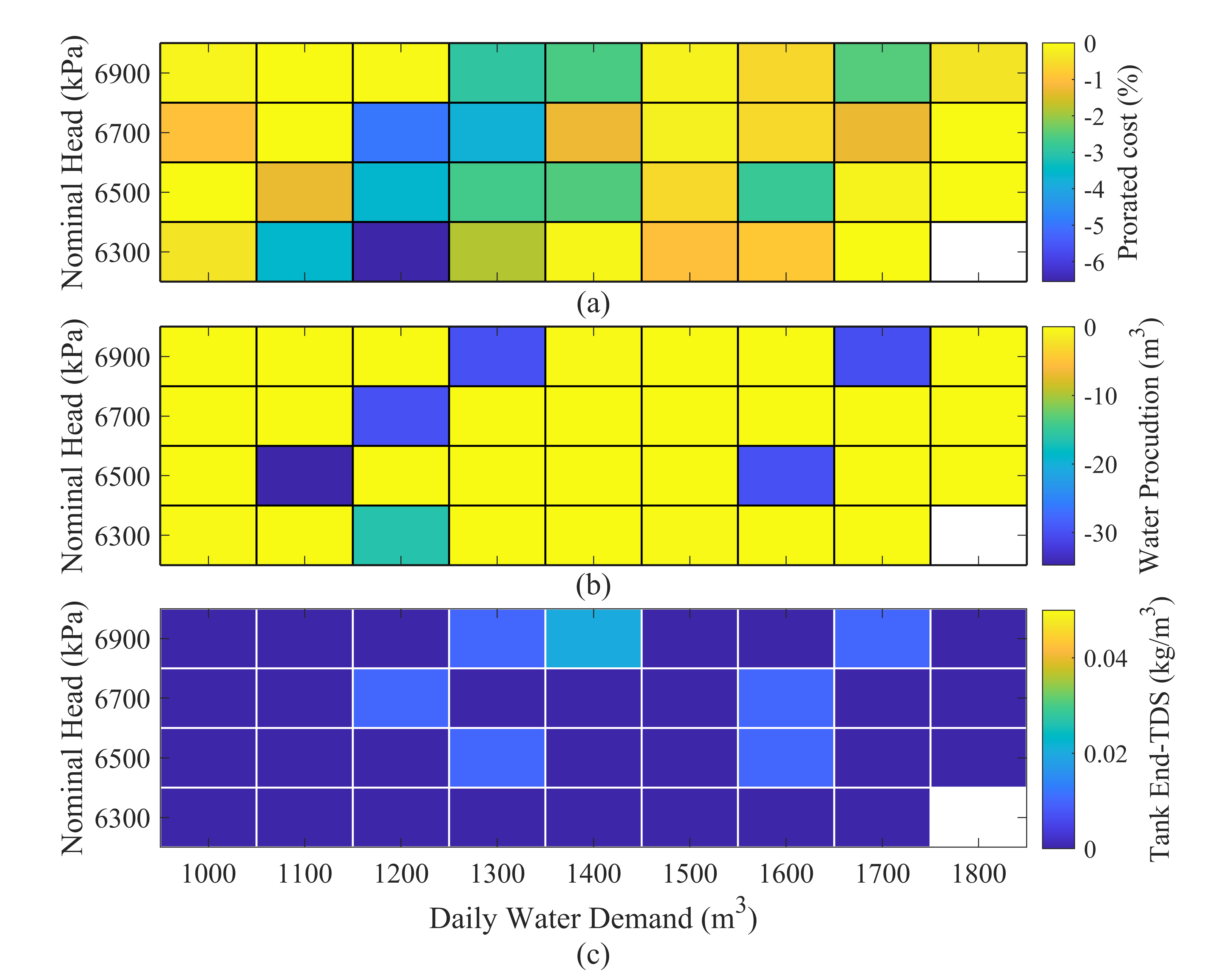}
    \caption{Sensitivity analysis of water demand and nominal head (\case{MixFlexIni}--\case{MixIni}): (a) prorated cost, (b) water production, and (c) tank end-TDS. Blank entries denote infeasible cases.}
    \label{fig:case_sense}
    \vspace{-6pt}
\end{figure}

\vspace{-3mm}
\subsection{Sensitivity Analysis}
Since daily water demand varies throughout the year and the nominal head is selected by the HDP operator, a sensitivity analysis of these two factors is conducted to evaluate the advantages of the proposed method in exploiting HDP flexibility. As shown in Fig.~\ref{fig:case_sense}(a), \case{MixFlexIni} achieves over 1\% cost reduction in 21 of the 36 operation scenarios, with a maximum reduction of 6.55\% at a nominal head of 6300~kPa and water demand of 1200~m$^3$/day. For all nominal-head settings, cost reduction first increases with water demand and then decreases. This occurs because the HDP operates mainly during low-price or high-PV periods when demand is very low, while limited flexibility remains at very high demand, as the system must run continuously at high production levels. Moreover, a higher nominal head increases water generation capability, shifting the demand range in which flexibility is beneficial to the right in Fig.~\ref{fig:case_sense}(a). Enabling flexibility from the tank TDS allows \case{MixFlexIni} to generate less additional water than \case{MixIni} in six scenarios, as shown in Fig.~\ref{fig:case_sense}(b), while the tank end-TDS of both cases remains almost the same.

\vspace{-2mm}
\subsection{Uncertainty}
\vspace{-1mm}
We generate 2,000 joint PV-electricity price scenarios and reduce them to 10 representative scenarios using the $k$-medoids method for scheduling. With the proposed TDCSO (Algorithm~\ref{alg:tdcso}), Step~1 solves $\boldsymbol{\mathcal{P}}^{\mathrm{SO\text{-}1}}_{\mathrm{SPLN}}$ using strategy \case{MixIni} to determine the commitment decisions in 161.6~s with an expected cost of \$621.0. For comparison, we also tested \case{MixFlexIni} in Step~1; it is substantially more time-consuming (629~s) and yields a lower expected cost of \$611.4, yet produces the same commitment decisions as \case{MixIni}. In Step~2, problem $\boldsymbol{\mathcal{P}}^{\mathrm{SO\text{-}2}}_{\mathrm{SPLN}}$ employs \case{MixFlexIni} and achieves an expected cost of \$609.48 under uncertainty in 652.3~s via parallel computation. Overall, solving the proposed TDCSO $\boldsymbol{\mathcal{P}^{\text{SO}}_{\text{SPLN}}}$ takes 813.9~s. As expected, incorporating uncertainties in scheduling results in a higher cost (+5.4\%) than the deterministic \case{MixFlexIni} case in Table~\ref{tab:ems_cmp}, with approximately tripled computation time. We also attempted to directly embed uncertainties into a single-step stochastic optimization under \case{MixFlexIni}; however, it cannot take advantage of parallel computation and failed to return a solution within 1~hour. These results demonstrate the effectiveness of the proposed TDCSO.

\section{Conclusion}
This paper proposes a coordinated operation strategy for power systems and RO-based HDP, fully exploiting the HDP's responsive flexibility. To reflect practical HDP operation, detailed models of the HPP, RO desalination process, and freshwater storage tank are developed, capturing key operational characteristics and requirements, including on-off operation with flushing, water quality constraints, and TDS mixing in the storage tank. Operational flexibility is further enhanced through proactive management of water storage and TDS mixing. The problem is appropriately simplified and linearized, as validated against the full desalination model. PV and electricity-price uncertainties are effectively incorporated through the proposed TDCSO approach. Case studies demonstrate up to 6\% reduction in operating cost while guaranteeing water production and end-user water quality. Future work will focus on large-scale desalination plants at the transmission level.



\bibliographystyle{IEEEtran}
\bibliography{Ref_desali_flex}

\vfill

\end{document}